\documentclass[fleqn,usenatbib]{mnras}
\usepackage{graphicx}
\usepackage{amssymb}
\usepackage{amsmath}
\usepackage{mathabx}
\usepackage{txfonts}
\usepackage[T1]{fontenc}
\usepackage{ae,aecompl}
\usepackage{cleveref}
\usepackage[super]{nth}

\usepackage[dvipsnames]{xcolor}

\title[High-order SD method in Low Mach flows]
{Spectral Difference method with a posteriori limiting: II- Application to low Mach number flows}

\author[D. A. Velasco-Romero and R. Teyssier]{%
David A. Velasco Romero$^{1,2}$\thanks{E-mail: david.velasco@ics.uzh.ch}, Romain Teyssier$^{1}$\\
$^{1}$Institute for Computational Science, University of Zurich, Winterthurerstrasse 190, 8057 Zurich, Switzerland\\
$^{2}$Department of Astrophysical Sciences, Princeton University,
4 Ivy Lane,
 Princeton, New Jersey 08544, United States.
}

\date{Accepted XXX. Received YYY; in original form ZZZ}

\pubyear{2022}

\begin{document}
\label{firstpage}
\pagerange{\pageref{firstpage}--\pageref{lastpage}}
\maketitle

\begin{abstract}
Stellar convection poses two main gargantuan challenges for astrophysical fluid solvers: low-Mach number flows and minuscule perturbations over steeply stratified hydrostatic equilibria. Most methods exhibit excessive numerical diffusion and are unable to capture the correct solution due to large truncation errors. In this paper, we analyze the performance of the Spectral Difference (SD) method under these extreme conditions using an arbitrarily high-order shock capturing scheme with a posteriori limiting. We include both a modification to the HLLC Riemann solver adapted to low Mach number flows (L-HLLC) and a well-balanced scheme to properly evolve perturbations over steep equilibrium solutions. We evaluate the performance of our method using a series of test tailored specifically for stellar convection. We observe that our high-order SD method is capable of dealing with very subsonic flows without necessarily using the modified Riemann solver. We find however that the well-balanced framework is unavoidable if one wants to capture accurately small amplitude convective and acoustic modes. Analyzing the temporal and spatial evolution of the turbulent kinetic energy, we show that our fourth-order SD scheme seems to emerge as an optimal variant to solve this difficult numerical problem.  
\end{abstract}


\defcitealias{velasco2023spectral}{Paper~I}
\newcommand\paperI{\citetalias{velasco2023spectral}}


\section{Introduction}
\label{sec:introduction}
Astrophysical fluid flows are characterized by a large range of conditions, from highly compressible and supersonic in the interstellar medium of galaxies, to highly subsonic deep inside stars. In general, specific numerical methods have to be developed for each limiting case. Finite volume (FV) or Finite Element (FE) time-explicit shock-capturing schemes are quite efficient in the supersonic regime \citep{FLASH,Teyssier2007, Dumbser2014,Schaal2015,cholla,Guillet2019,WENO-WOMBAT,Athena++,fambri2020discontinuous,2023MNRAS.522..982C}, while spectral anelastic solvers \citep{1997JCoPh.131...89R,2006JCoPh.219...21B,2013ApJ...772...21R,2015JCoPh.299..374S}, based on the anaelastic approximation \citep{1969JAtS...26..448G}, or FV time-implicit schemes \citep{MUSIC,2013PhDThesis..M,2017A&A...600A...7G,2021A&A...653A..55H,2023MNRAS.519.5333B} are traditionally used in the subsonic regime. Spectral anelastic schemes entirely remove sound waves by filtering them out of the solution, while FV time-implicit solvers usually suffer from convergence and parallelization issues. In order to explore intermediate regimes (sound waves within low Mach number flows) or to increase the computational efficiency, it has been proposed in recent years to extend the capabilities of finite volume schemes to the conditions of low Mach number flows \citep{Maestro,2013ApJ...773..137G,2015A&A...576A..50M,Mastroex}. 

The low Mach number regime is particularly challenging for time-explicit FV schemes because the time step has to decrease linearly with the sound speed owing to the Courant-Friedrichs-Lewy (CFL) stability condition. For low Mach number flow,  a prohibitively large number of time-steps are required to reach a given convective time scale. This causes truncation errors (a.k.a numerical diffusion) to accumulate over time, smearing out interesting details in the solution. Two solutions have been proposed to solve this issue: high-order methods \citep{nicoud2000conservative,desjardins2008high, klein2016high, Apsara, 2024arXiv240216706L} and low Mach number Riemann solvers \citep{2015A&A...576A..50M,2016arXiv161203910B}. In particular, high-order FV or FE methods exponentially reduce numerical diffusion (the so-called exponential convergence property), drastically reducing the impact of the cumulative error budget after many time-steps.

In this paper, we would like to explore the performance of the Spectral Difference (SD) method in the low Mach number regime. The SD method has proven particularly interesting because it lies at the boundary of FE and FV schemes. This interesting property allows us to develop hybrid strategies to control spurious oscillations in presence of discontinuities \citep{Dumbser2014,Vilar2019,2021JCoPh.42609935H,2022FrP....10.8028C,velasco2023spectral}. Our recent implementation of the SD method \citep[][hereafter Paper~I]{velasco2023spectral} is based on a posteriori limiting to prevent oscillations and to preserve positivity. In this new paper, we show that high-order SD is a promising method to model low Mach number flows. We show good performance compared to previous high-order FV implementation or low Mach number Riemann solvers. We also apply our scheme to the poster child example of astrophysical low Mach number flow, namely stellar convection \citep{2006ApJ...642.1057H}. 

In \paperI, our arbitrarily high-order implementation of the SD method was presented. We designed shock-capturing capabilities via {\it a-posteriori} limiting, in which a robust second-order method is used as fallback scheme. The fallback scheme is used to recompute and replace high-order fluxes whenever the solution to be updated is consider to be inadmissible. As demonstrated in \paperI, the SD method is strictly equivalent to a Finite Volume (FV) method using sub-element control volumes. This strict equivalence allows the use of a standard second-order FV method as the fallback scheme responsible of providing robustness to the method, namely a positive and non-oscillatory solution. The second-order method chosen as the fallback of our implementation is no other than the classical MUSCL-Hancock scheme (Monotonic Upstream-centered Scheme for Conservation Laws) described in \citet{1979JCoPh..32..101V}. 

\paperI~presents a series of tests to assess the performance of our method, both for the advection equation and for the Euler equations. The results showed an overall benefit in the solution when going to higher-order, even in highly supersonic problems with strong shocks. The present paper is a follow-up of this previous work, extending the study to highly subsonic flows, relevant to stellar convection. 

Stellar convection, in addition to its low Mach number nature, brings a particularly severe additional challenge due to the presence of a stratified medium in hydrostatic equilibrium.  Convection is characterized by very small density and temperature perturbations around a strictly static equilibrium solution. In order to properly evolve these small perturbations of the reference equilibrium state, we have developed a so-called well-balanced scheme  \citep{1996SIAMJ.33...1G,veiga2019capturing,2021A&A...652A..53E} specifically for the SD method. The simple idea behind the design of well-balanced schemes is to evolve the perturbations rather than the complete solution, still using non-linear fluxes via the Riemann solver. Provided that the equilibrium state is known in advance, this scheme can greatly enhance the quality of the solution, even for classical second-order schemes \citep{2007JCoPh.226...29N, 2014ApJ...786...24H,  2016A&A...587A..94K, veiga2019capturing, 2021A&A...653A..55H,2020A&A...641A..18H,2021A&A...652A..53E, cuissa2022toward}. Note that well-balanced schemes have been often designed precisely for stellar convection \citep{2021A&A...653A..55H,2020A&A...641A..18H,2021A&A...652A..53E, cuissa2022toward}. 

For this work, we have developed a new code using the \texttt{python} language and the {\ttfamily cupy} package to enable the use of Graphical Processing Units (GPU). It is available upon request to the authors. The paper is divided as follows: In \autoref{sec:method}, we present the original numerical method. In \autoref{sec:results}, we present the results of our systematic study to illustrate the performance of SD in the low Mach number regime, as well as in the context of stellar convection. In \autoref{sec:discussion}, we discuss our results obtained, and finally in \autoref{sec:conclusions}, we draw our conclusions.

\section{Numerical method}
\label{sec:method}
\subsection{Governing equations}
\label{sec:ge}
The Euler equations in vector form write in 2D and in Cartesian coordinates as:
\begin{equation}
    \partial_t \mathbf{U} + \partial_x \mathbf{F}(\mathbf{U}) + \partial_y \mathbf{G}(\mathbf{U}) = \mathbf{S}(\mathbf{U})
\end{equation}
where the solution vector $\mathbf{U}$ and the flux vectors $\mathbf{F}$ and $\mathbf{G}$ are:
\begin{equation}
\mathbf{U} = 
    \begin{pmatrix}
    \rho \\
    \rho v_x \\
    \rho v_y \\
    E
    \end{pmatrix}, \quad
\mathbf{F} = 
    \begin{pmatrix}
    \rho v_x \\
    \rho  v^2_x + P \\
    \rho v_x v_y \\
    (E + P)v_x
    \end{pmatrix},  \quad
\mathbf{G} = 
    \begin{pmatrix}
    \rho v_y \\
    \rho v_x v_y \\
    \rho  v^2_y + P \\
    (E + P)v_y
    \end{pmatrix}. 
\end{equation}
where $\rho$ is the density of the fluid, $\mathbf{v} =(v_x,v_y)$ is the velocity field, $E = e + \frac{1}{2}\rho |v|^2$ the total energy density, and $P$ the pressure. The system of equations is closed with the equation of state of an ideal gas $P = (\gamma-1)e$, where $\gamma$ is the adiabatic index.
The source term we consider here is only due to the gravitational acceleration $\mathbf{g} = (g_x, g_y)$:
\begin{equation}
\mathbf{S}(\mathbf{U}) = 
    \begin{pmatrix}
    0 \\
    \rho g_x \\
    \rho g_y \\
    \rho (g_x v_x + g_y v_y)
    \end{pmatrix}
\end{equation}

\subsection{Well-Balanced Scheme}
\label{sec:wb}
Near equilibrium solutions consist of small perturbations over a known equilibrium solution:
\begin{equation}
    \mathbf{U} = \mathbf{U}_\text{eq} + \mathbf{U}',
\end{equation}
where the equilibrium solution satisfies
\begin{equation}
    \partial_x \mathbf{F}(\mathbf{U}_\text{eq}) + \partial_y \mathbf{G}(\mathbf{U_\text{eq}}) = \mathbf{S}(\mathbf{U}_\text{eq}).
\end{equation}
Well-balanced schemes evolve the perturbations rather than the complete solution, taking advantage of the fact that $\partial_t\mathbf{U} = \partial_t\mathbf{U}'$,
since $\partial_t\mathbf{U}_\text{eq}=0$. 
Interpolation to interface values is performed over both perturbation and equilibrium solution, so that the complete solution at the left and right interfaces is:
\begin{equation}
    \mathbf{U}_{i\pm 1/2} = \mathbf{U}_{\text{eq},i\pm 1/2} + \mathbf{U}_{i}' \pm \partial_x\mathbf{U}_{i}'\Delta x_i/2.
\end{equation}
The Riemann problem at each interfaces is solved using the left and right complete solutions as:
\begin{equation}
\hat{{\bf F}}_{i\pm\frac{1}{2}}  = {\bf F}\left( RP \left[ {{\bf U}}_{i\pm1/2}^{L},{{\bf U}}_{i\pm1/2}^{R}\right] \right).
\end{equation}
A perturbation flux is simply defined by:
\begin{equation}
    \hat{{\bf F}}'  = \hat{{\bf F}} - \hat{{\bf F}}_\text{eq}.
\end{equation}
The perturbation (rather than the solution) is finally updated using:
\begin{equation}
    \mathbf{U}^{'n+1}_{i} = \mathbf{U}^{'n}_{i} 
    -\left(\frac{\hat{\bf{F}}^{'n+1/2}_{i+1/2}-\hat{\bf{F}}^{'n+1/2}_{i-1/2}}{h_{i}}\
    \right)\Delta t.
\end{equation}
This method prevents the numerical solution (small perturbations to the equilibrium solution) from being dwarfed by large truncation errors in the equilibrium solution. This general framework can be extended easily to the SD methodology we present now.

\subsection{Spectral Difference Method}
Here we present a summary of the method described in \citetalias{velasco2023spectral}.
First of all, the computational domain is decomposed into non-overlapping elements, explaining why fundamentally SD is a FE method. Inside each element, a continuous high-order numerical solution $\mathbf{U}(x)$ is given by:
\begin{equation}
    \mathbf{U}(x) = \sum_{m=0}^p \mathbf{U}(x^s_{a,m})\ell^s_m(x),
    \label{eq:sol}
\end{equation}
where $\{\ell^s_{m}(x)\}_{m=0}^p$ represents the set of Lagrange interpolation polynomials up to degree $p$ built on the set of $p+1$ solution points $\mathcal{S}^s=\{x^s_{a,m}\}_{m=0}^p$ in element $a$.
A second set of Lagrange polynomials up to degree $p+1$ on a set of $p+2$ flux points $\mathcal{S}^f=\{x^f_{a,m}\}_{m=0}^{p+1}$ in element $a$, is used to represent the high-order approximation of the flux:
\begin{equation}
\mathbf{F}(x) =  \sum_{m=0}^{p+1} \mathbf{F}(x^f_{a,m}) \ell^f_m(x).
\label{eq:flux}
\end{equation}
We define the numerical fluxes as:
\begin{equation}
\hat{\mathbf{F}}(x) =  \sum_{m=0}^{p+1} \hat{\mathbf{F}}(\mathbf{U}(x^f_{a,m})) \ell^f_{m}(x),
\end{equation}
where $\hat{\mathbf{F}}(\cdot)$ denotes the numerical flux resulting from solving the Riemann problem at the interface between elements. 
The solution is then updated as:
\begin{align}
\label{eq:semi-discrete}
\frac{{\rm d}}{{\rm d}t} \mathbf{U}(x^s_{a,m},t) = {\cal L}_{a,m}(\mathbf{U})= -\sum_{m'=0}^{p+1} \hat{\mathbf{F}}(\mathbf{U}(x^f_{a,m'},t)) \ell^{\prime f}_{m'}(x^s_{a,m}).
\end{align}
where $\ell^{\prime}$ is the derivative of the Lagrange polynomials.

The fully discrete method is achieved when combining the SD spatial discretization with a time integration discretization. In this work we pair SD with the so-called ADER time-integrator, so that the resulting fully discrete method can be written as:
\begin{equation}
\mathbf{U}(t + \Delta t) = \mathbf{U}(t) + \Delta t \sum_{k=0}^p w_{k} {\cal L}(\mathbf{U}^{p}_k).
\label{eq:sd_update}
\end{equation}
where $k$ iterates over the $p+1$ time slices of the ADER time-integrator (for more details see \paperI). 

As shown in \paperI, the SD method is equivalent to the FV method at the sub-element, control volume level, allowing for the fully-discrete scheme to be also written as:
\begin{equation}
\bar{\mathbf{U}}_{a,m}(t + \Delta t) = \bar{\mathbf{U}}_{a,m}(t) - \Delta t \sum_{k=0}^p w_k\frac{\left(\hat{\mathbf{F}}^{f,k}_{a,m+1}-\hat{\mathbf{F}}^{f,k}_{a,m}\right)}{h_{m}},
\end{equation}
where $\bar{\mathbf{U}}_{a,m}$ is the control-volume-averaged solution and $h_m$ is the size of the control volume within each element.

\subsection{A Posteriori Limiting}
As described in \paperI, we use a second order Godunov scheme \citep{1979JCoPh..32..101V} as a fallback scheme when either spurious  oscillations or non physical values are observed in the high-order candidate solution. We identify troubled cells for each control volume. This is a key difference of our scheme compared to other previous implementations. Limiting is performed within elements, at the level of each individual sub-element control volume.
The control volume fluxes pertaining the troubled sub-cells are recomputed with our monotonicity- and positivity-preserving fallback scheme (namely the MUSCL-Hancock scheme, hereafter FV2). We then recompute the control volume averages (at each ADER stage) of all the subcells that share these recomputed fluxes: 
\begin{equation}
    \bar{\mathbf{U}}^{k+1}_{i,j} = \bar{\mathbf{U}}^k_{i,j} - {\left(\frac{\hat{\mathbf{F}}^{k}_{i+1/2,j}-\hat{\mathbf{F}}^{k}_{i-1/2,j}}{h_i} + \frac{\hat{\mathbf{F}}^{k}_{i,j+1/2}-\hat{\mathbf{F}}^{k}_{i,j-1/2}}{h_j}\right)}w_k\Delta t,
\end{equation}
where here $\hat{\mathbf{F}}$ represents a combination of high-order and fallback scheme fluxes.

\subsubsection{Low Mach number Riemann solver}
The HLLC Riemann solver \citep{HLLC} is known to exhibit excessive numerical diffusion at low Mach numbers, failing to deliver the correct incompressible limit \citep{LHLLC}. \citet{minoshima2021low} present a modification to the HLLD Riemann solver (that can be trivially extended to HLLC) that allows for the proper behaviour at low Mach numbers. This improvement of the original HLLC is based on a correction for the contact wave pressure ($P_*$). It starts with the computation of the dimensionless value $\chi$:
\begin{equation}
    \chi = \min\left( 1, \frac{\max(|v_L|,|v_R|)}{c_{max}}\right),
\end{equation}
where $c_{\max} = \max(c_L,c_R)$ is the maximum between the left and the right sound speed. The coefficient that reduces the diffusion term in the low Mach number limit is :
\begin{equation}
    \phi   = \chi(2-\chi).
\end{equation}
where the corrected contact wave pressure is:
\begin{align*}
    P_* &= \frac{\rho_R(s_R-v_R)P_L + \rho_L(v_L-s_L)P_R}{\rho_R(s_R-v_R)+\rho_L(v_L-s_L)}\\
    &+ \frac{\phi\rho_R(s_R-v_R)\rho_L(v_L-s_L)(v_L-v_R))}{\rho_R(s_R-v_R)+\rho_L(v_L-s_L)},
\end{align*}
and where $s_{L,R}$ are the left and right wave speeds:
\begin{align}
    s_L = \min(v_L,v_R)-c_{\max},\quad
    s_R = \max(v_L,v_R)+c_{\max}.
\end{align}
It is important to stress that the resulting low-Mach number scheme requires a significantly reduced CFL. In theory, the time-step should be reduced by an additional factor proportional to the Mach number. In practice, we observe that the time step reduction does not need to be as large but it can still be significant (see the following sections).

\subsubsection{Flux Blending}
As mentioned before, troubled sub-cells are updated by replacing high-order (SD) fluxes with low-order (FV2) ones. In doing so, the neighboring sub-cells, in case of not being troubled themselves, are updated with a combination of SD and FV2 fluxes, as the flux pertaining to the interface shared with the troubled sub-cell is replaced with a FV2 flux. This procedure, depicted in Figure~3 of \paperI, does not guarantee admissibility for the corrected solution at trouble-adjacent sub-cells.
A brute-force solution to this, would be to iterate the detection procedure until the solution of all sub-cells is admissible.
This procedure, although robust, can result to be too computationally expensive.

\begin{figure}
    \centering
    \includegraphics[width=\columnwidth]{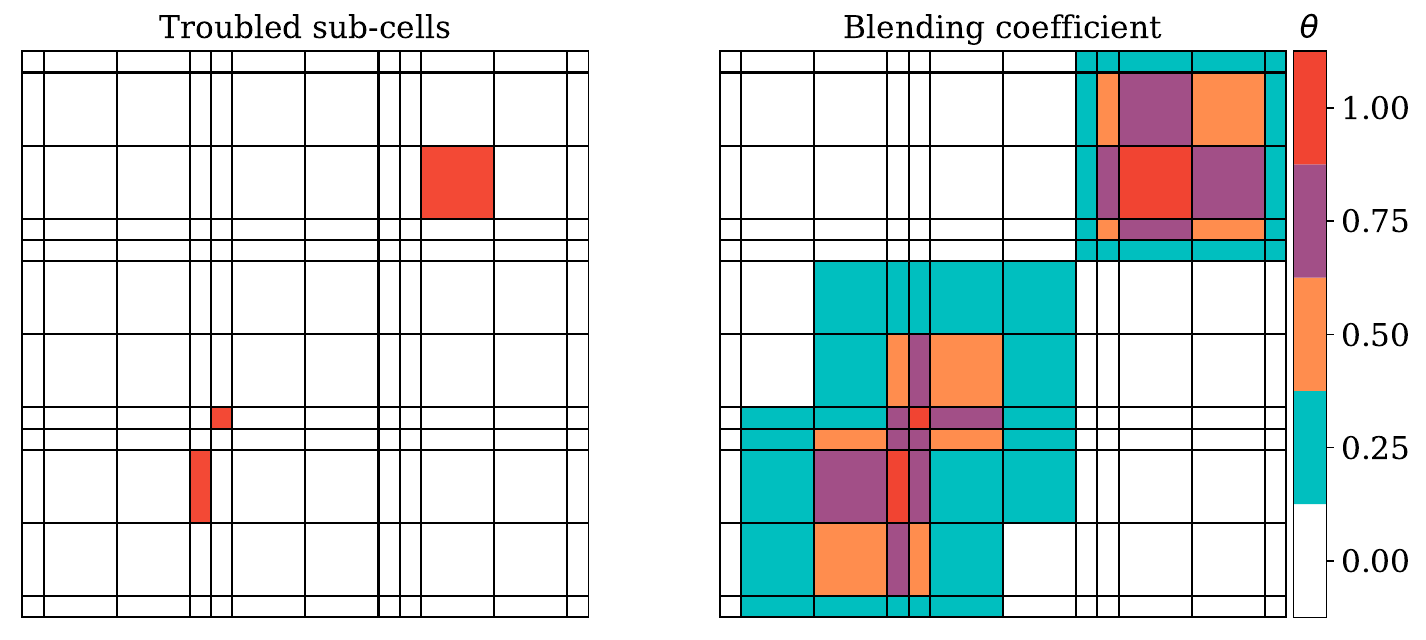}
    \caption{Flux blending: On the left in red the troubled sub-cells. On the right the values for the blending coefficient $\theta$}
    \label{fig:blending}
\end{figure}

As described in \citet{2021JCoPh.42609935H,vilar2024}, a less expensive solution, although not perfectly robust, is to blend the SD and FV2 fluxes at trouble-adjacent sub-cells in a convex combination.

\begin{equation}
    \hat{\mathbf{F}}_{i\pm\frac{1}{2},j} = (1-\theta_{i\pm\frac{1}{2},j})\mathbf{F}^{\text{SD}}_{i\pm\frac{1}{2},j} + \theta_{i\pm\frac{1}{2},j} \mathbf{F}^{\text{FV}}_{i\pm\frac{1}{2},j}
\end{equation}

where $\theta_{i-\frac{1}{2},j}= \max(\theta_{i-1,j},\theta_{i,j})$ is the blending coefficient.
This coefficient has value $\theta_{i,j}=\{1,3/4,1/2,1/4,0\}$ at troubled sub-cells, at first, second and third neighbors, and elsewhere respectively. Considering sub-cell $(i,j)$ to be troubled, $\theta$ takes the following values:
\begin{align*} 
\theta = 
\begin{cases}
     1 \quad & (i,j) \\
     3/4 \quad &(i\pm1,j), (i,j\pm1)\\
     1/2 \quad &(i\pm1,j\pm1) \\
     1/4 \quad &(i\pm2,j), (i\pm2,j\pm1),(i\pm2,j\pm2),\\ &(i\pm1,j\pm2), (i,j\pm2) \\
     0 \quad &\text{otherwise}
    \end{cases}
\end{align*}
As shown in \autoref{fig:blending}, for sub-cells neighboring multiple troubled sub-cells, $\theta$ takes the value assigned by the nearest troubled sub-cell.

\subsubsection{Limiting criteria}
The limiting criteria, used to determine troubled sub-cells, consist of a Numerical Admissibility Detection (NAD) and a Physical Admissibility Detection (PAD), as described in \paperI~and similar to  \citet{Vilar2019}.

The NAD criteria requires for the candidate solution to be bounded by the local extrema at the previous time-step:
\begin{equation}
  \min(\bar{\mathbf{U}}_{i-1}^{k},\bar{\mathbf{U}}_{i}^{k},\bar{\mathbf{U}}_{i+1}^{k}) \leq \bar{\mathbf{U}}_{i}^{k+1} \leq \max(\bar{\mathbf{U}}_{i-1}^{k},\bar{\mathbf{U}}_{i}^{k},\bar{\mathbf{U}}_{i+1}^{k}),  
  \label{eq:nad}
\end{equation}
These criteria results to stringent, as it cannot distinguish between discontinuous and smooth extrema, where smooth extrema are considered features of the solution rather than spurious artifacts of the method. In order to relax these criteria and allow smooth extrema to evolve unlimited, we add a Smooth Extrema Detection (SED), where at least the first numerical derivative of the candidate solution is required to be continuous (\paperI).

The PAD criteria requires for the candidate density and pressure to be above a minimum value. For this work we use $\rho_{\min}=10^{-10}$ and $P_{\min}=10^{-10}$.

\subsubsection{Well-Balanced Spectral Difference Scheme}
For the well-balanced property of the method the NAD criteria is applied on the perturbation rather than on the solution \citep[see more details in][]{veiga2019capturing}. On the other hand, the PAD criteria needs to be apply on the full solution in order to ensure positivity for density $\rho$ and pressure $P$.

\section{Numerical results}
\label{sec:results}

In this section we present a series of tests that demonstrate the good performance of our method when describing low-Mach number flows, both for smooth and non-smooth solutions. 
We also test our method when small perturbations are advanced in time on top of a steep equilibrium solution. We compare the performance of our well-balanced scheme to the high-order SD solution without a well-balance scheme. 
Our final test is meant to include all these different aspects in a more complex and astrophysical relevant scenario, namely the evolution of a buoyantly rising bubble in an hydrostatic stellar atmosphere. 

In this paper, we define the number of degrees of freedom as $N_{\rm DOF}=N_x(p+1) \times N_y(p+1)$, which is the product of the $(p+1)$ solution points per dimension in each element by $N_x$ (and $N_y$), the number of elements in the x (and y) direction.
We also recall that $p$ is the polynomial degree of our solution within each element, and that $(p+1)$ is also the order of accuracy of the resulting numerical approximation. The adiabatic index, for every test, was set to a value of $\gamma=5/3$.

In this section, we will compare our different Spectral Difference methods to the well-known second-order MUSCL-Hancock methods. We will use the straightforward notations SD3, SD4, etc for Spectral Difference of order 3, 4, etc and FV2 for MUSCL-Hancock, as in Finite Volume (FV) second-order. The limiting in MUSCL-Hancock is performed using the \textit{moncen} slope limiter \citep{van1977towards}, to provide the least possible numerical diffusion for this method.

In the context of low Mach number flows, it is also important to consider the low Mach number fix of the HLLC Riemann solver, called L-HLLC, proposed by \citet{LHLLC,minoshima2021low} to minimize numerical diffusion. For SD, we use L-HLLC only in the flux calculation of the fallback scheme. We name the corresponding schemss SD3L, SD4L, etc. We use the name FV2L to refer to the MUSCL-Hancock scheme paired with L-HLLC.

For SD, we also have the option to use a convex combination of high-order and second-order fluxes, at trouble-adjacent sub-cells, to minimize the discontinuous switch to the fallback scheme. In case we turn on blending, we call the corresponding scheme SD3B, SD4B, etc. In summary, the scheme called SD4BL corresponds to the 4th-order spectral difference method ($p=3$) with L-HLLC for the fallback scheme and flux blending.

We will see in the following section that better quality solutions can be obtained for low Mach number flows if one combines the higher-order accuracy of SD with the low diffusivity of L-HLLC.

\subsection{Gresho vortex}
\label{sec:gresho}

The Gresho vortex test \citep{gresho1990theory} has been originally designed to test the performance of high-order FV and FE scheme in conserving angular momentum and preserving properly this stationary solution \citep{2015A&A...576A..50M,Apsara,Velasco2018}. This test is particularly challenging for methods using a Cartesian mesh. This test can also be used to test the performance of the method in the presence of highly subsonic conditions, using the setup we present below. 

The Gresho vortex test is defined by the following initial conditions for the fluid density, x- and y-velocity and pressure:
\begin{align*} 
\rho &= 1, \quad v_x = -v_\phi y_c/r +v_0, \quad v_y = v_\phi x_c/r,\\
P &= \frac{1}{\gamma \mathcal{M}^2_{\rm max}}-\frac{1}{2} +
\begin{cases}
     \frac{25}{2}r^2 \, &r<0.2 \\
     4\log(5r) + 4 -20r + \frac{25}{2}r^2 \, &0.2<r<0.4\\
     4\log(2) - 2 \, &r>0.4,
    \end{cases}
\end{align*}
where $x_c = x-0.5$, $y_c = y-0.5$,  $r=\sqrt{x_c^2+y_c^2}$. $v_0=5$ is the background velocity and $v_\phi$ is the azimuthal velocity with:
\begin{align*} 
v_\phi = 
\begin{cases}
     5r \quad &r<0.2 \\
     2-5r \quad &0.2<r<0.4\\
     0 \quad &r>0.4
    \end{cases}
\end{align*}
in a box with dimensions $x,y \in [0,1]$ with $N_x = N_y = N$ elements, periodic boundary conditions and an adiabatic index $\gamma=5/3$. In this test we refer to the Mach number as $\mathcal{M} = v_{\phi}/c_s$, where for the setup at hand, the maximum value for the Mach number is $\mathcal{M}_{\max} = 10^{-1},10^{-2},10^{-3}$, found at $r=0.2$. The simulation is run until time $t=1$ which corresponds to 5 advection cycles of the vortex across the periodic box.

\begin{figure}
    \centering
    \includegraphics[width=0.85\columnwidth]{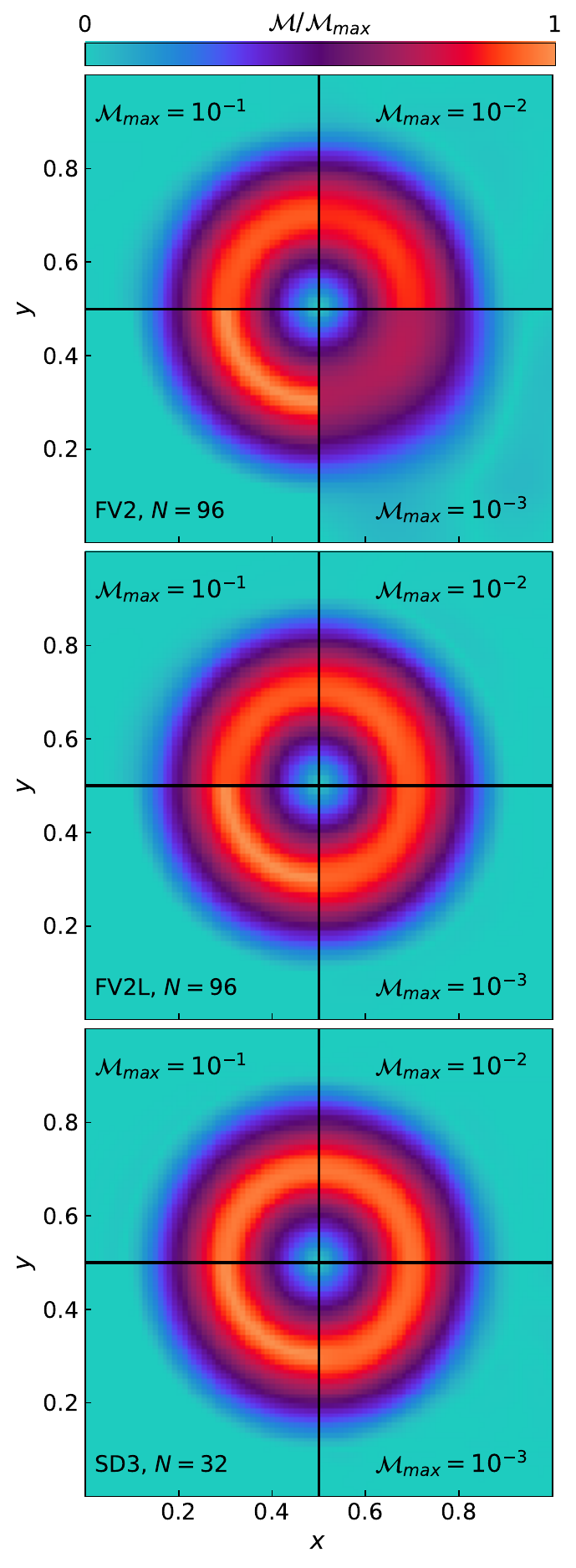}
    \caption{Gresho vortex test: Color maps for the control volume average of $\mathcal{M} = v_{\phi}/c_s$. Results at $t = 1$ for three values of the Mach number ($\mathcal{M}_{\max}=10^{-1},10^{-2}$ and $10^{-3}$) and a background velocity $v_0=5$. On the first row the results for FV2 (with HLLC), on the second row FV2L (with L-HLLC), and and on the third row SD3 ($p=2$). All of them making use of $96^2$ DOF.}
    \label{fig:gresho}
\end{figure}

In \autoref{fig:gresho} we present the results of the method with $N_\text{DOF}=96^2$ for FV2 (the original MUSCL-Hancock scheme with HLLC Riemann solver), FV2L (MUSCL-Hancock with L-HLLC) and SD3 (Spectral Difference scheme with $p=2$ or third order). Each panel shows 4 quadrants with a portion of the image of the local Mach number, normalized by the maximum value. The lower left quadrant shows the initial condition that should be preserved for this stationary smooth solution. The other quadrants show the final solution after one revolution for different Mach numbers. We see that FV2 suffers from increased numerical diffusion owing to the increased sound speed. Using L-HLLC greatly improves the quality of the solution, even though the peak of the Mach number distribution has been smoothed out. Going to SD3, we see that the higher order of accuracy really helps in preserving the initial solution. It is now less clear to see a deviation from the expected stationary solution. It is important to mention that while for FV2, SD3 and SD4, a Courant coefficient of $C=0.8$ was used at all Mach numbers, it was necessary to use $C=0.05$ for FV2L at the lowest Mach number case in order to recover stability.

\begin{figure}
    \centering
    \includegraphics[width=0.85\columnwidth]{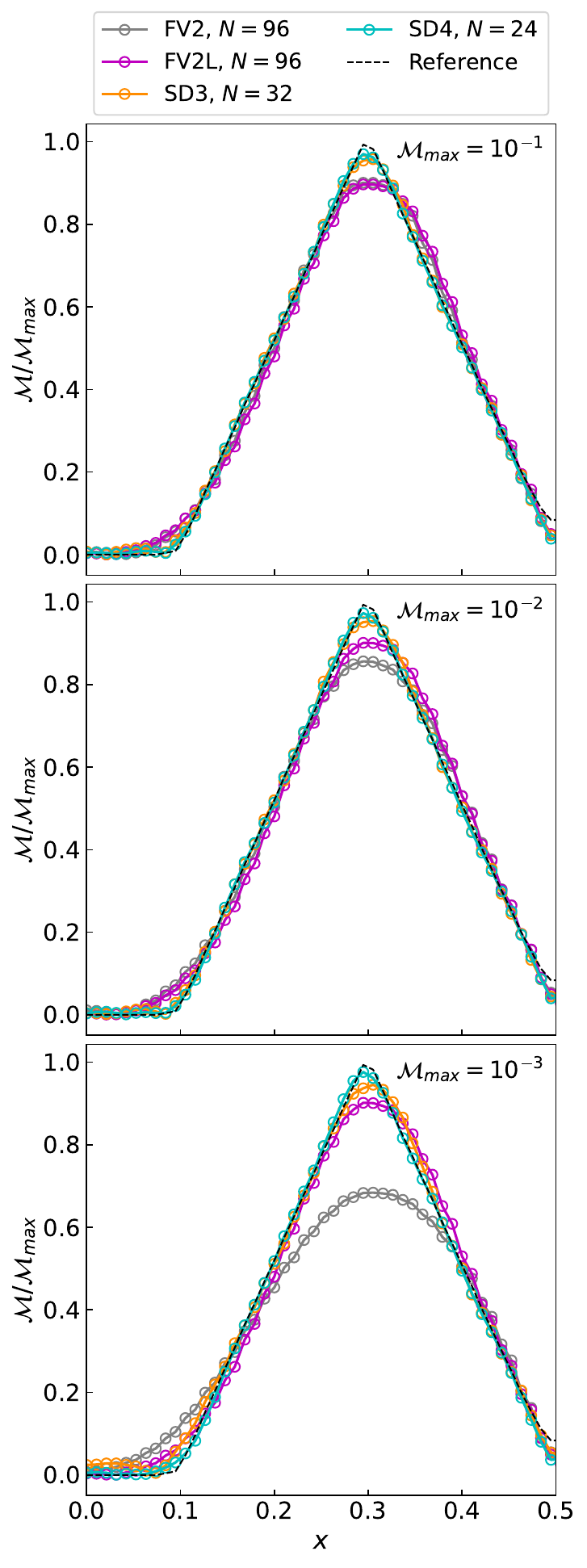}
    \caption{Gresho vortex test: 1-dimensional slices of the control volume average of $\mathcal{M} = v_{\phi}/c_s$ at $t=1$, corresponding to five laps over the domain ($v_0=5$).}
    \label{fig:gresho2}
\end{figure}

In \autoref{fig:gresho2} we present one-dimensional cuts at $y=0.5$. Each panel shows our results for a different value of the maximum Mach number. Note that all simulations presented here use the same $N_\text{DOF}$, namely $96$ per dimension. FV2 is very sensitive to the increased sound speed. As a result, the amplitude of the vortex has decayed by almost a factor of 3 during the 5 laps. On the other hand, FV2L does not show any visible dependence on the Mach number anymore, although the numerical solution is still being affected by the numerical diffusion of this second-order scheme. Finally, SD3 and SD4 both show a very good preservation of the initial profile, with increasing quality with increasing order of accuracy. Note that in this case, because the profiles are so smooth, our SD methods never trigger the fallback scheme.

\subsection{Rayleigh-Taylor instability}

The Rayleigh-Taylor instability (RTI) is the basic phenomenon that occurs in a stratified, convectively unstable medium. It features a sharp interface between two fluids of different density. With gravity pointing upwards, the lighter fluid is placed on top the denser fluid, giving rise to an interchange instability, where the light fluid rises buoyantly and the heavy fluid sinks \citep{chandrasekhar2013hydrodynamic}. Secondary Kevin-Helmholtz instabilities usually form due to shearing motions between rising and sinking flows. 

This test is particularly challenging for high-order methods as it is fundamentally not smooth. Indeed, the discontinuity between the 2 fluids will activate most slope limiters and degrade locally the order of accuracy to lower order. 

It is customary to describe the effect of numerical diffusion using the modified equation analysis, resulting in the following approximate Taylor expansion for the numerical scheme:
\begin{equation}
\frac{\partial U}{\partial t}+ \frac{\partial F}{\partial x} \propto \left( |v| + c_s\right) h^{p+1} \frac{\partial^{p+2} U}{\partial x^{p+2}} \simeq \nu_{\rm num} \frac{\partial^{2} U}{\partial x^2}
\end{equation}
where the leading error term has been reformulated here using a classical diffusion operator. This rough derivation allows us to define a numerical diffusion coefficient
\begin{equation}
    \nu_{\rm num} \simeq \left( |v| + c_s\right) h \left(\frac{h}{L}\right)^{p}
    \label{eq:numdiff}
\end{equation}
even though for high-order schemes the exact nature of numerical errors can be significantly more complex. We can see clearly the adverse effect of the sound speed that increases the numerical diffusion compared to the contribution of the velocity field alone. We can also see the effect of the mesh size and of the order of the method. 

In the previous expression, $L$ is the typical length scale of the numerical solution. For well-resolved, smooth features, high-order will clearly help in reducing the numerical diffusion. For sharp discontinuities for which $L \simeq h$, this is not the case anymore.  This simple formalism will help us interpret our results in the current section and in the ones to follow.

The setup for this test is a slight modification of the setup described in \citet{liska2003comparison}. The initial conditions are as follows:
\begin{align*}
    \rho, P &= 
    \begin{cases}
         \rho_1, P_0+\rho_1 y \quad &y\leq y_c\\
         \rho_2, P_0+\rho_2 y + (\rho_1-\rho_2)y_c \quad &y>y_c 
    \end{cases}
    \\
    v_x &= 0, \quad v_y = -0.025 \Delta v \cos(8\pi x)
\end{align*}
where $y_c=0.5$, $\rho_1=2$, $\rho_2=1$ and $\gamma=5/3$. The amplitude of the velocity perturbation is set by $\Delta v=\sqrt{\gamma(P_c-P_0+1)/\rho_1}$. The domain of the simulation is a box with dimensions $x,y \in [0,0.25]\times[0,1]$ with $N_x=N/4$ and $N_y=N$ elements. We use periodic boundary conditions in $x$  and reflective boundary conditions in $y$.

This test is interesting because we can change the midplane pressure $P_0$ without changing the velocity perturbation $\Delta v$. The corresponding Mach number will be inversely proportional to the midplane sound speed $c_s=\sqrt{\gamma P_c/\rho_1}$. In principle, we expect our solution to converge to the incompressible solution as ${\cal M} \rightarrow 0$ or $c_s \rightarrow +\infty$.

\begin{figure}
    \centering
    \includegraphics[width=.98\columnwidth]{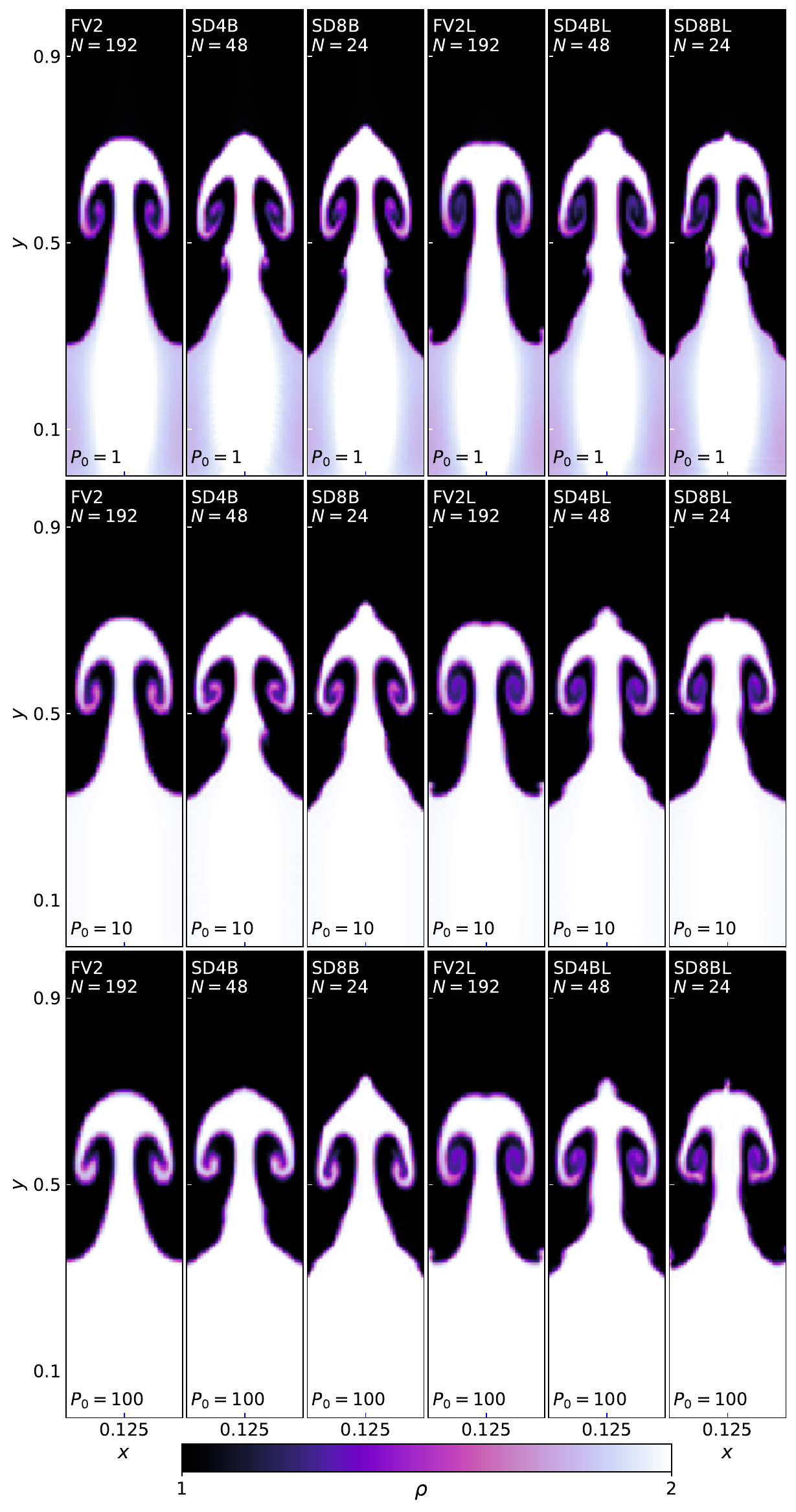}
    \caption{Rayleigh Taylor instability test. Colormaps for the density at $t=1.95$ for \nth{2}, \nth{4} and \nth{8} order of the numerical approximation for $48\times192$ DOF. On the first row the results for $P_0=1$, on the second row the results for $P_0=10$ and on the third row the results for $P_0=100$.}
    \label{fig:rt}
\end{figure}

\begin{figure}
    \centering
    \includegraphics[width=.98\columnwidth]{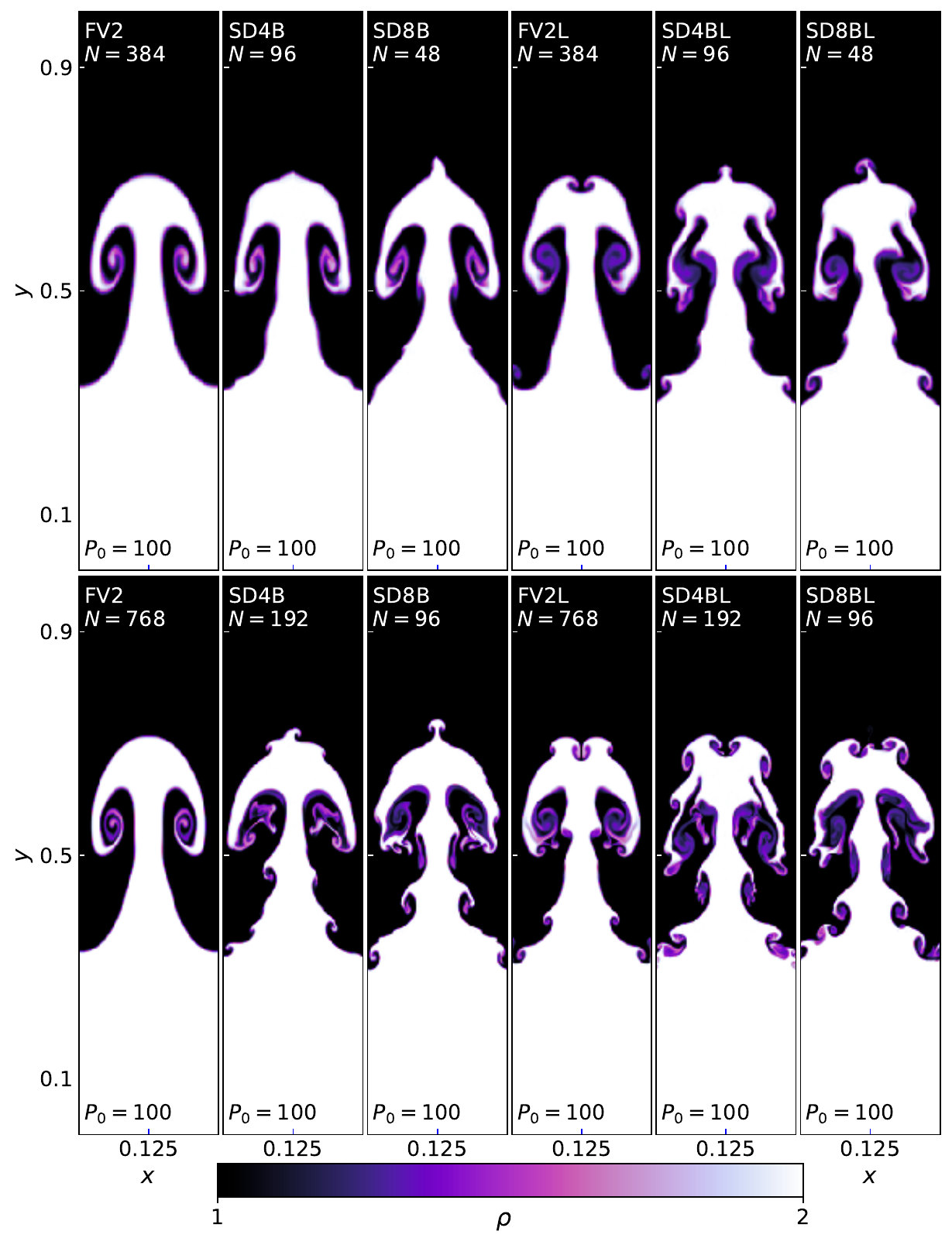}
    \caption{Rayleigh Taylor instability test. Iso-contours for the density at $t=1$ for  \nth{2}, \nth{4} and \nth{8} order of the numerical approximation for $P_0=100$. On the first row the results with $96\times384$ DOF, on the second row the results with $192\times768$ DOF.}
    \label{fig:rt2}
\end{figure}

In \autoref{fig:rt} we present the results of our different numerical schemes for this test. Each row corresponds to a different value of $P_0=$1, 10 and 100. The left column shows the results for FV2 and $N=192$ cells in the vertical dimension. We clearly see that the solution is not converging towards the incompressible solution. It does quite the opposite, with more and more numerical diffusion with increasing sound speed, in agreement with Equation~\ref{eq:numdiff}. 

If we use instead a low-Mach number Riemann solver (scheme FV2L, fourth column in Fig.~\ref{fig:rt}), we recover the desired result, namely that the solution is now independent of the Mach number and consistent with the incompressible limit. Indeed, as explained in \citet{minoshima2021low}, the low Mach number fix has been designed to precisely remove the sound speed dependence in the numerical flux, allowing us by construction to recover the proper low-Mach number limit. 

We now adopt our high-order SD scheme at 4th and 8th order, using flux blending to limit the effect of our discontinuous fallback strategy. Results are visible in the second and third column of Fig.~\ref{fig:rt}. We see that higher-order accuracy helps in preserving the expected solution, as predicted by Equation~\ref{eq:numdiff}. Finally, using the L-HLLC Riemann solver for the fallback scheme in conjunction with higher-order polynomial reconstruction gives us the best of both worlds, with a solution that is independent of the Mach number but that also shows more non-linear features (SD4BL and SD8BL in the last two columns of Fig.~\ref{fig:rt}). 

Figure~\ref{fig:rt2} shows the effect of increasing $N_\text{DOF}$ to 384 and 768 (2x and 4x compared to Fig.~\ref{fig:rt}) only for the highest sound speed case $P_0=100$. We see that for FV2, the increased resolution barely compensates for the effect of the increased sound speed, again in agreement with Equation~\ref{eq:numdiff}. If one uses L-HLLC instead of HLLC, we see much more non-linear details emerging with better resolution. Note that our results are in striking agreement with \citet{LHLLC} for this particular Rayleigh-Taylor problem and our fiducial second-order MUSCL-Hancock scheme. 

Going to high-order shows that the numerical diffusion is even more reduced, allowing us to see more and more non-linear and chaotic features in the solution. This supports the claim that the effective numerical Reynolds number increases with increased order, especially if our high-order SD schemes are used in conjunction with L-HLLC for the fallback scheme. Indeed, SD4BL and SD8BL show the most prominent non-linear features, even though they used the same number of DOF than FV2 and FV2L.   

\subsection{Acoustic Perturbation of Hydrostatic Equilibrium}

The test \citep{xing2013high} describes a non-trivial hydrostatic equilibrium solution, altered by a tiny perturbation in pressure, described in this section as ``the acoustic pulse''. The unperturbed hydrostatic solution we consider here is given by:
\begin{align*}
  \rho_{\text{eq}} &= \rho_0\exp\left(-\frac{\rho_0}{P_0}\phi\right), \quad v_{x,\text{eq}} = 0, \quad v_{y\text{eq}} = 0,\\
  P_{\text{eq}} &= P_0\exp\left(-\frac{\rho_0}{P_0}\phi\right)
\end{align*}
where $\rho_0=1$, $P_0=1$ and $\phi(x,y) = g(x+y)$ is the gravitational potential corresponding to an oblique gravitational acceleration  with $g_x=g_y=g=1$. 
We then add in the initial condition a small circular pressure perturbation:
 \begin{align*}
\delta P = P - P_{\text{eq}} = \eta\exp\left(-\frac{\rho_0 g}{P_0} \frac{r_p}{0.01}\right)
\end{align*}
where $r_p = (x-0.3)^2 + (y-0.3)^2$ is the distance to the center of the acoustic pulse and $\eta$ is the amplitude of the pulse. The computational domain is a box of dimensions $[0,1]\times[0,1]$ with $N_x = N_y = N$ elements in each direction. For boundary conditions, we impose in the ghost elements the unperturbed hydrostatic profile. Finally, the adiabatic index for this test was set to $\gamma=5/3$.

\begin{figure}
    \centering
    \includegraphics[width=0.85\columnwidth]{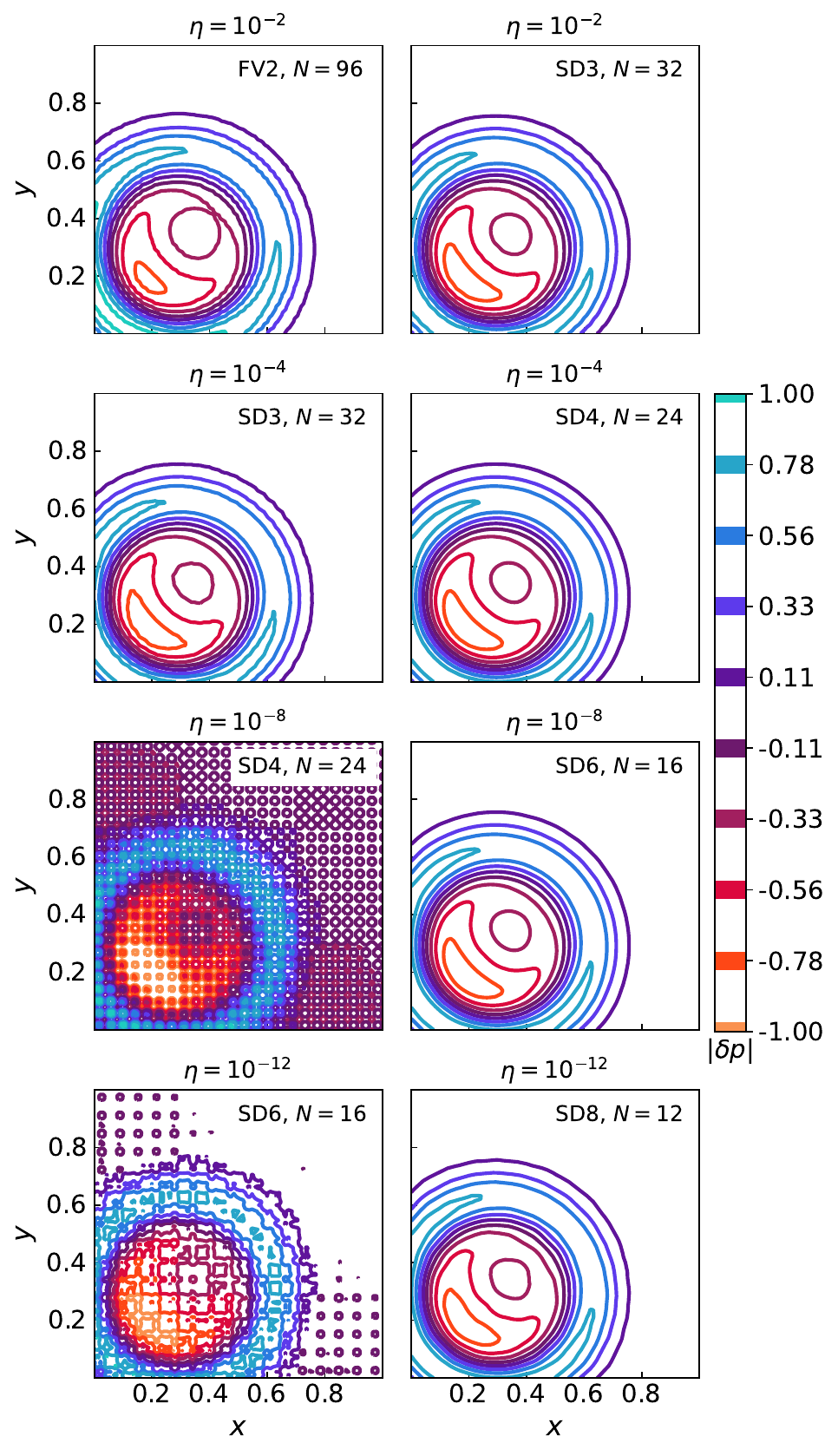}
    \caption{Perturbation over hydrostatic equilibrium test. Iso-contours at $t=0.25$ of $\delta P$ for different values of the perturbation amplitude ($\eta= 10^{-2}, 10^{-4}, 10^{-8}$ and $10^{-12}$), with a total of $96^2$ DOF.}
    \label{fig:hs_p}
\end{figure}

In \autoref{fig:hs_p} we show iso-contours of the pressure perturbation $\delta P$ at the final time $t=0.25$. Each row corresponds to a different value of the perturbation amplitude 
$\eta= 10^{-2}, 10^{-4}, 10^{-8}$ and $10^{-12}$. We use a total of $96^2$ DOF for FV2, SD3 ($p=2$), SD4 ($p=3$), SD6 ($p=5$) and SD8 ($p=7$). As usual, we have reduced the number of elements for higher order polynomial reconstructions to keep the number of DOF constant. Note that in this first test, we don't use our well-balanced scheme. 

In the first row we can observe that even for $\eta= 10^{-2}$, the largest perturbation in this study, FV2 fails at recovering the right solution. Truncation errors in the equilibrium solution are already strong enough to alter the accuracy of the acoustic pulse solution. We need the third-order accuracy of SD3 to get an accurate solution. In the second row we see that SD4 does a very good job for $\eta= 10^{-4}$, but fails at reproducing the correct solution with $\eta= 10^{-8}$ (a perturbation four orders of magnitude smaller). 

In the last row, it is apparent that SD8 is able to properly recover the acoustic pulse, even with $\eta= 10^{-12}$, a perturbation amplitude close to machine precision. Note however that in this extreme case, the iso-contours exhibit some small oscillations, showing that our SD8 scheme is probably reaching its limits. 
It is remarkable that a second-order scheme completely fails to describe a $\eta= 10^{-2}$ perturbation, while  our eighth-order scheme is capable of describing a perturbation eight orders of magnitude smaller {\it using the same number of DOF}.  We have verified that our solution with FV2 fails at recovering the correct solution with $\eta= 10^{-8}$ even using $3072^2$ DOF ($32^2\times$ more elements that the present test).

\begin{figure}
    \centering
    \includegraphics[width=0.9\columnwidth]{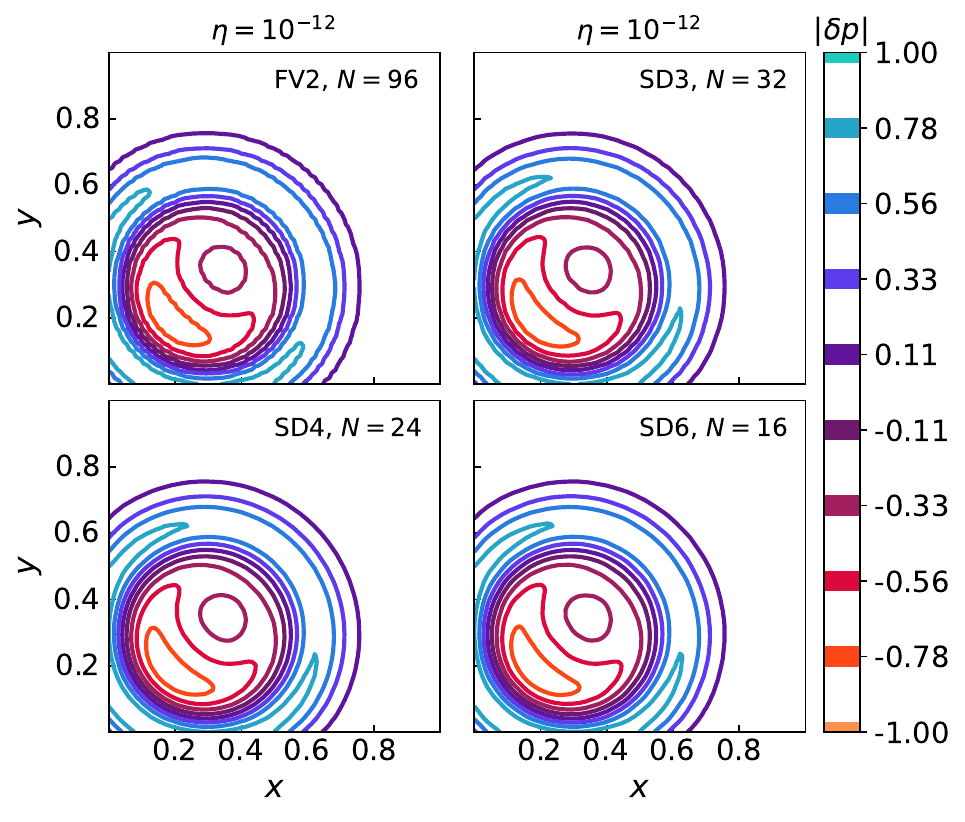}
    \caption{Small perturbation of hydrostatic equilibrium test: Results using FV and SD with a well-balanced scheme. 
    Iso-contours at $t=0.25$ of $\delta P$ with initial perturbation amplitude $\eta=10^{-12}$, with $96^2$ DOF for the \nth{2}-, \nth{3}, \nth{4}- and \nth{6}-order.}
    \label{fig:hs_wb}
\end{figure}

We repeat now this test making use of our well-balanced scheme (see \autoref{sec:wb}). \autoref{fig:hs_wb} shows the results obtained for FV2, SD3, SD4 and SD6 with the smallest amplitude $\eta=10^{-12}$. It is clear that the well-balanced property of the scheme allows us to recover in this case the right solution, even at second-order with as few as $96^2$ DOF. The expert eye could however see that the contours are better defined when using higher order schemes. 

\subsection{Turbulent Convection}

This final integrated test of our new well-balanced SD scheme features a buoyantly rising bubble in a convectively unstable hydrostatic atmosphere that transitions into decaying convective turbulence. This test combines 4 particularly challenging requirements for our code, namely 1- a very low Mach number turbulent flow, 2- a steep hydrostatic profile, 3- discontinuous features, and last but not least, 4- very long time integration. This test is quite relevant for studying turbulent convection in stars and its interaction with nearby radiatively stable layers. In particular, if one uses a compressible hydro solver, one can study the interplay between turbulence and internal gravity waves, with many applications in astro-seismology \citep{2013ApJ...772...21R,2023ApJS..264...21B}. 

\begin{figure*}
    \centering
    \includegraphics[width=0.95\textwidth]{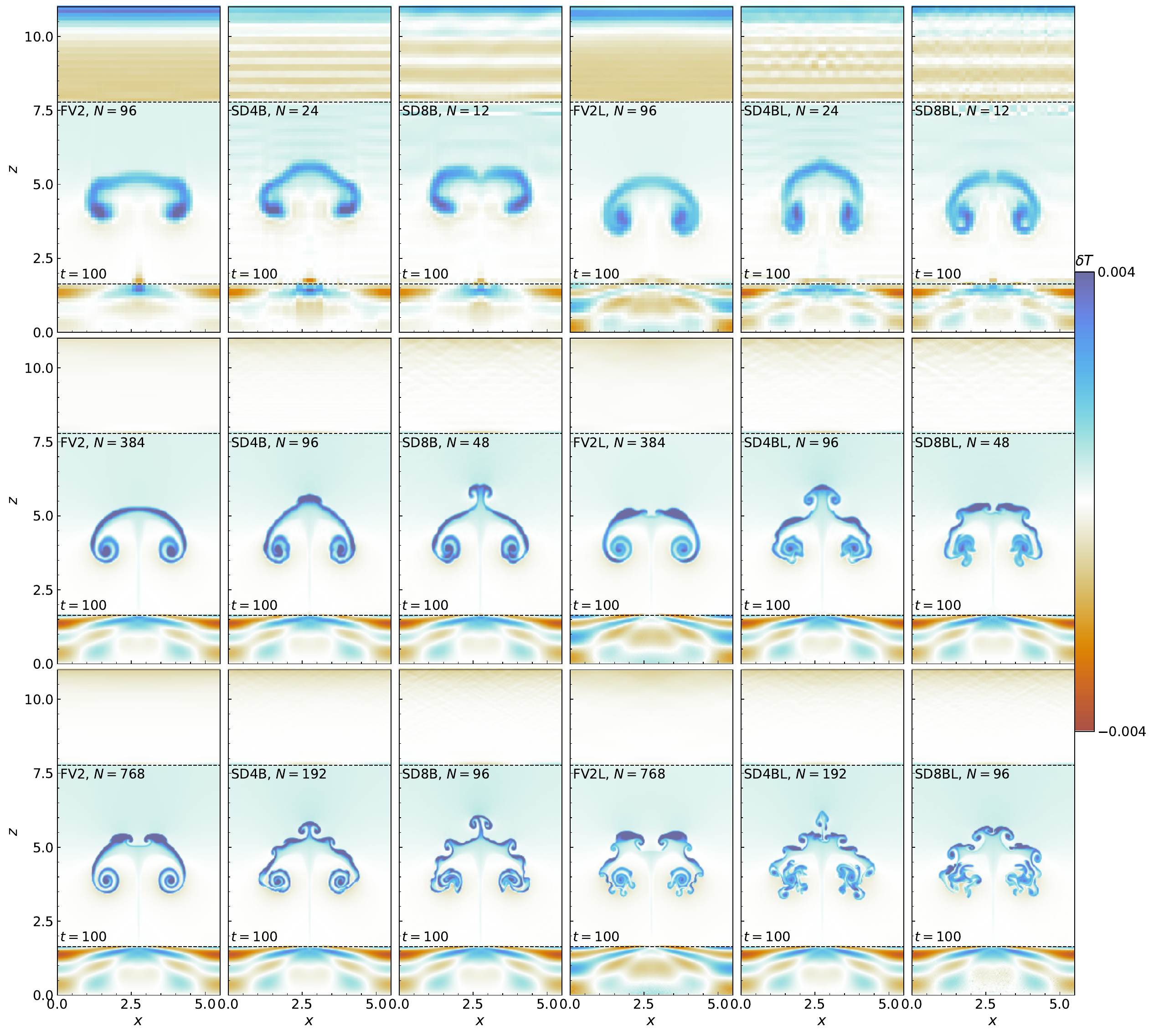}
    \caption{Temperature perturbation for the buoyantly rising bubble test at $t=100$. Results for FV2, SD4B and SD8B with and without L-HLLC.}
    \label{fig:bubbles}
\end{figure*}

In order to test our numerical techniques, we consider a simplified setup with a 2D square box with a constant gravity vector pointing downward. Our initial conditions are inspired by previous works on the study of the He flash in AGB stars \citep{herwig2004evolution, 2006ApJ...642.1057H} and on the study of the properties of convective dynamos \citep{cuissa2022toward}. The simulation domain is a tall 2D rectangular box with $x,z \in [0, L/2]\times[0,L]$. The size of the box in the z-direction is set in code units to be $L = 11$. The domain is divided vertically into 3 separate regions (or zones) described by 3 hydrostatic equilibrium solutions based on 3 different polytropic indices. These 3 equilibrium solutions are defined by the following constants:
\begin{align*}
    \rho_0,P_0,\gamma_0 \simeq 
    \begin{cases}
        1.81, 5.94, 1.20 \quad &z_0 \leq z\leq z_1 (\text{radiative zone}),\\
        1.17, 3.53, 1.67 \quad &z_1 \leq z\leq z_2(\text{convective zone}),\\
        0.09, 0.05, 1.01 \quad &z_2 \leq z\leq z_3 (\text{radiative zone}),\\
    \end{cases}
\end{align*}
where $z_0=0$, $z_1=1.64$ and $z_2=7.77$ and $z_3=L$. The values of $\rho_0$ and $P_0$  corresponds to the density and pressure of the fluid at the bottom of each region, while $\gamma_0$ is the polytropic index in each region. 
The temperature profile in each region $i=1,2,3$ is described by the following linear profile: 
\begin{equation}
\frac{T}{T_0} = 1-\frac{(\gamma_0-1)}{\gamma_0}\frac{\rho_0 g}{P_0} \left( z-z_{i-1}\right).
\end{equation}
where the gravitational acceleration is set to $g_z=-g$ with $g=1$. 
The density and pressure profiles follow the polytropic relations:
\begin{equation}
\rho = \rho_0 \left( \frac{T}{T_0}\right)^{\frac{1}{\gamma_0-1}}~{\rm and}~
P = P_0 \left( \frac{T}{T_0}\right)^{\frac{\gamma_0}{\gamma_0-1}}.
\end{equation}
Note that all these profiles are continuous across each  boundary but not differentiable. 
If we were to use these equilibrium profiles, using our well-balanced SD scheme, nothing would happen as hydrostatic equilibrium is strictly preserved. 

We then introduce at $t=0$ a small perturbation in the density using a circular bubble defined as:
\begin{align*}
    \frac{\delta \rho}{\rho_{\rm eq}} = 
    \begin{cases}
    -0.012 \exp{\left(-\frac{r^2}{2r_b^2}\right)}, \quad & r < r_b, \\
    0, \quad & \text{otherwise,}
    \end{cases}
\end{align*}
The bubble radius is $r_b=0.25$. The center of the bubble is located at $x_b = L/4$, $z_b = z_1+0.75$. Note that the bubble is underdense so that it immediately accelerates and rises buoyantly in the hydrostatic atmosphere. The value of the adiabatic index is once again $\gamma=5/3$.

\begin{figure*}
    \centering
    \includegraphics[width=0.95\textwidth]{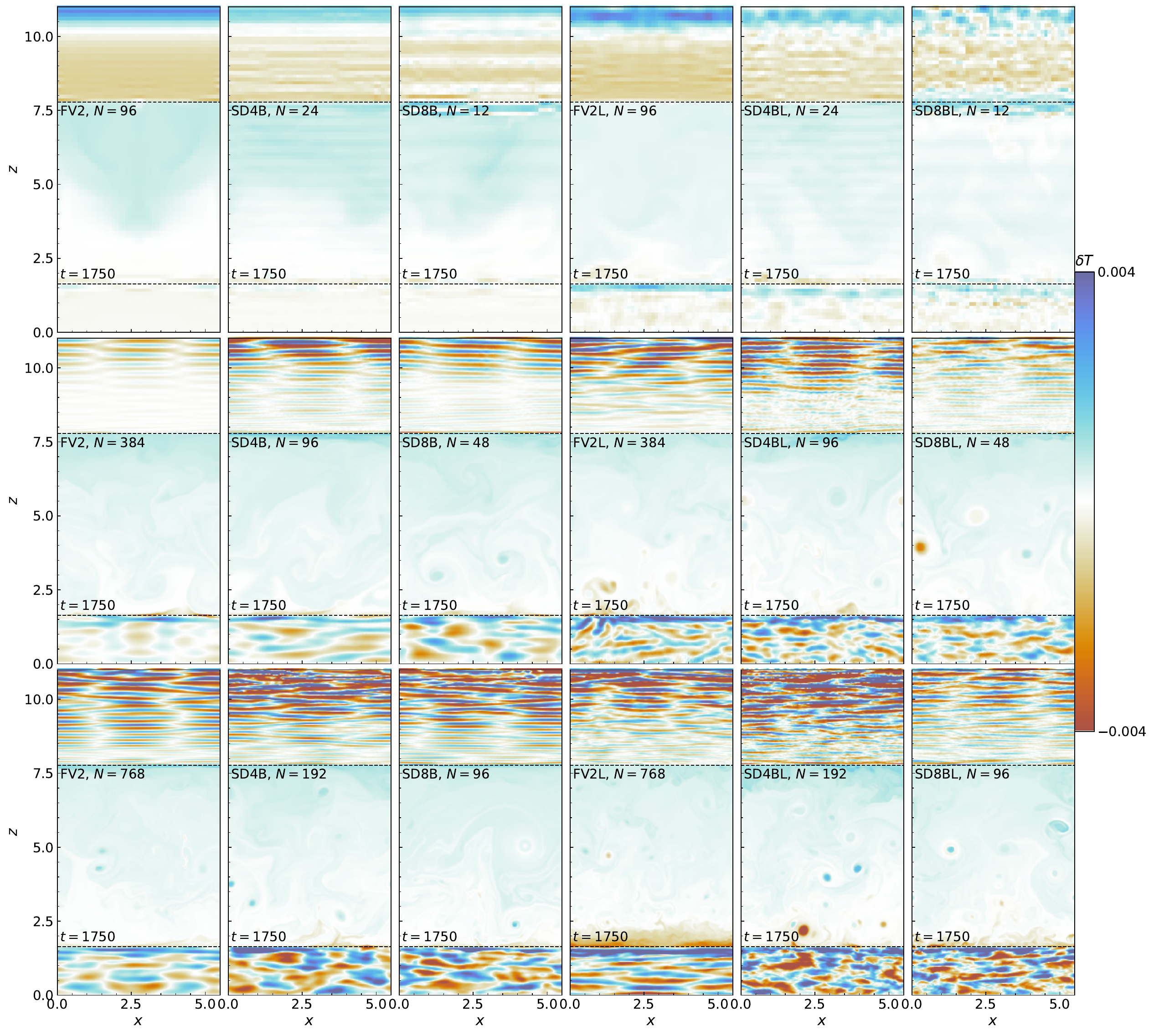}
    \caption{Temperature perturbation for the buoyantly rising bubble test at $t=1750$. Results for FV2, SD4B and SD8B with and without L-HLLC.}
    \label{fig:bubbles_late}
\end{figure*}

\begin{figure*}
    \centering
    \includegraphics[width=0.95\textwidth]{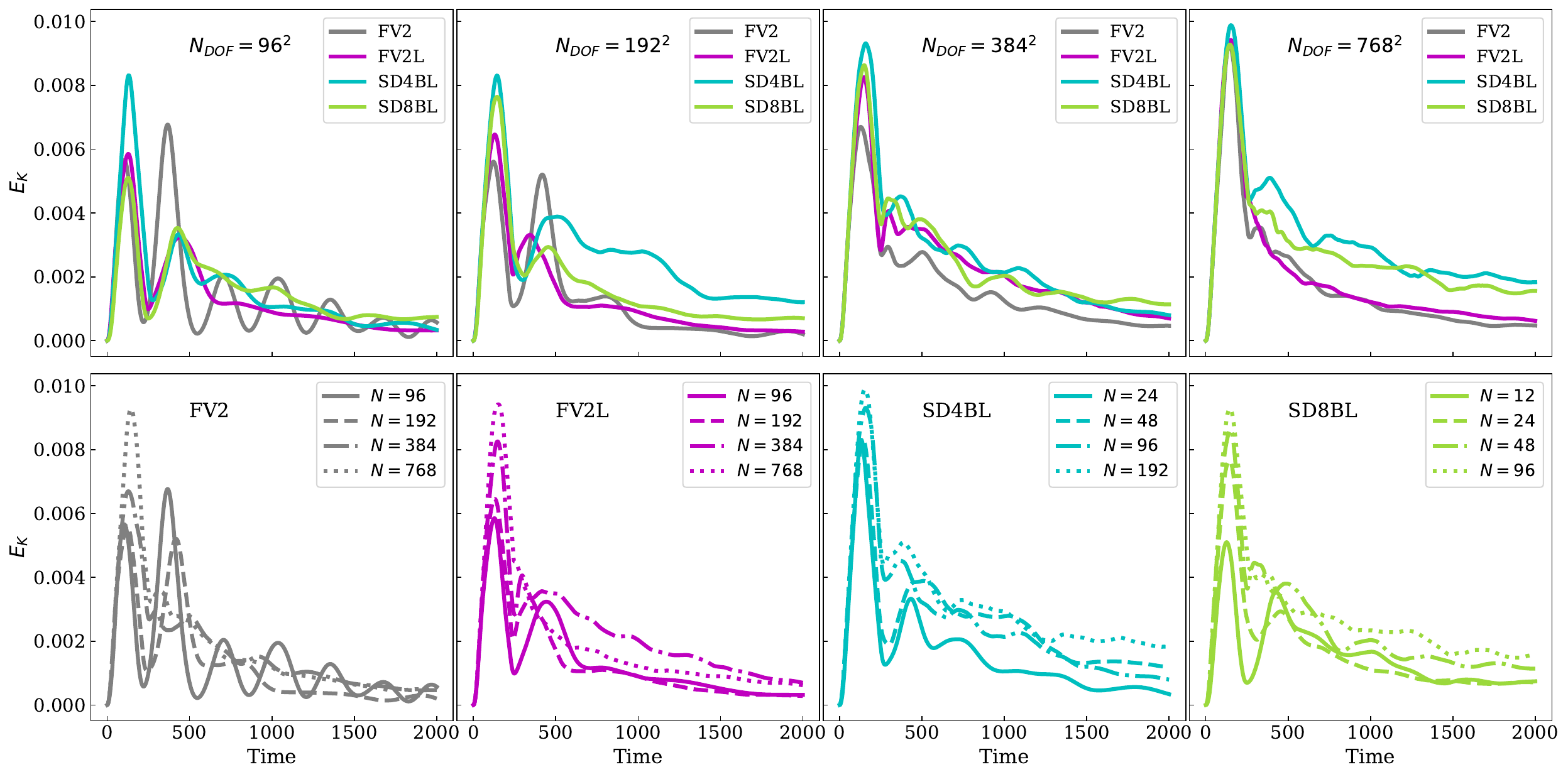}
    \caption{Time evolution for the total kinetic energy of the buoyant bubble test. The top row shows (from left to right) our different schemes using similar number of DOF. The bottom row on the other hand, shows the effect of increasing the number of DOF for a given numerical scheme.}
    \label{fig:Ek_t}
\end{figure*}

In \autoref{fig:bubbles}, we present maps of the perturbation in the gas temperature $\delta T/T_{\rm eq}$ at $t=100$, corresponding to the peak of the kinetic energy of the bubble as seen in Figure~\ref{fig:Ek_t}. The top row show our results using only 96 DOF along the y-axis (the longer axis). Our traditional well-balanced second-order scheme without low Mach number fix to the HLLC Riemann solver (labelled FV2) shows the most numerical diffusion. Increasing the order of accuracy helps a little bit getting closer to the correct solution. SD4B and SD8B corresponds to our SD scheme with flux blending but again using HLLC for the fallback scheme. 

The most important factor here is the addition of the low Mach number fix via the L-HLLC Riemann solver. Indeed, at this low resolution, FV2L recovers the right overall bubble shape. As for the RTI test, going to high order helps getting slightly more non-linear features. Note our highest-order scheme, SD8BL, only needs 12 elements across the vertical axis, each element containing the 8 additional degrees of freedom corresponding to the 8 sub-element control-volumes. Note that the internal gravity wave is already visible on the lower stable region. Compared to the other panels, we see that the solution obtained with SD4BL and SD8BL seems already converged. 

The middle row corresponds to a 4x increase of resolution for each scheme. FV2 is now getting the correct bubble shape but it appears quite smooth. It corresponds in fact to the solution obtained by FV2L at a resolution 4x less. Using L-HLLC with FV2L helps in getting more prominent KH rollups. The most spectacular change comes from combining high-order and low-Mach number fix with SD4BL and SD8BL in the 2 right panels of the central row. We see a clear transition to turbulence with multiple secondary instabilities already breaking up the bubble. 

The bottom row shows the results of our 6 different schemes at 2x better resolution. FV2 $N=768$ is again very similar to FV2L $N=384$ even with $2^2\times$ more cells. We see now that the bubble obtained with FV2L is breaking up into multiple secondary instabilities, in agreement with what we obtained with SD4BL and SD8BL with a much lower number of DOF. At this much higher resolution, our high-order schemes using L-HLLC for the fallback scheme show the most prominent turbulent features, supporting our earlier claim that the effective numerical Reynolds number is much higher in this case. Note also that the waves in the lower stable zone are fully converged and quite smooth.

In \autoref{fig:bubbles_late}, we present maps of the temperature fluctuation at  $t=1750$ (lower panels). This corresponds to a time when turbulence has fully developed across the convective region and to a significant decay (one order of magnitude) of the peak kinetic energy. Our most diffusive scheme, FV2, shows a very laminar and symmetric flow structure at the lowest resolution, at the higher resolution (middle and bottom row left panels) where the flow appears more chaotic. In comparison, the high-order schemes develop chaotic features much earlier in terms of resolution. 

It is interesting to see how large amplitude compressive waves have now entered the top stable region. These acoustic waves are due to the bubble bouncing of the top boundary before falling back down and breaking up into multiple bubblets. It is striking to see that high-order schemes (and to a lesser extent FV2L) develop long lived vortices visible as the red and blue circular features in the convective zone. These vortices share many properties with the Gresho vortex that high-order methods are so superior at capturing. The vortices are totally absent in the FV2 solution, except when using our highest resolution setup. Note that these long-lived vortices are a well-known features of 2D decaying turbulence and would probably be much harder to see in 3D convective flows.

On larger scales, the convective region shows the characteristic convective features with cold plumes going down and hot bubbles going up. The velocity field is dominated by large scale perturbations, typical of subsonic 2D turbulence. Note that the waves in the top and bottom stable zones are not consistent anymore between the different orders of accuracy and seem to evolve in a rather chaotic way.

In \autoref{fig:Ek_t}, we show the time evolution of the kinetic energy in the convective zone using our different high-order schemes at different resolution. 
The top row compares different order of accuracy using the same number of DOF, while the bottom row compares the results from the same high-order scheme but at different resolution. We see a clear trend where going to higher order and going to higher resolution helps preserving the kinetic energy of the bubble. 

The maximum value of the kinetic energy is increasing and shows some clear signs of convergence as one increases the polynomial degree and the number of elements. After the first peak around $t=100$, the kinetic energy decays slowly as turbulence slowly dissipates. It is striking to see the kinetic energy almost entirely dissipated by $t=1000$ for FV2 (expect for the lowest resolution, which exhibits an oscillatory behaviour), while other schemes combining the low-Mach number L-HLLC Riemann solver and high-order polynomial reconstruction with $p=3$ or $p=7$ manage to maintain the late time kinetic energy much better. The very slow decay of kinetic energy at late time is consistent with very high effective Reynolds number. Our SD8BL scheme with the highest number of elements $N=96$ provides the lowest effective  numerical dissipation. 

It is striking to see much more oscillations in Figure~\ref{fig:Ek_t} for FV2 than for any of the other less dissipative schemes. This is consistent with the laminar nature of the flow in this case, with very symmetric features bouncing back and forth as the bubble dissipates buoyantly. The much more chaotic nature of the solution in the other cases modify the kinetic energy time evolution quite drastically, with only the first peak still visible, while the late time evolution only shows a secular slow decrease.  

If one uses only the kinetic energy as the figure of merit for these different schemes, one can see from Figure~\ref{fig:Ek_t} that using L-HLLC appears as the most important ingredient, while increasing the order of the scheme reduces dissipation even more. The question that immediately arises here is the question of the cost of SD4BL compared to FV2L. Indeed, one can see that doubling the resolution with FV2L can easily compensate for the extra cost associated to SD4B or SD4BL. The question of cost will be discussed in \autoref{app:perf}. One weakness of FV2L is the requirement to reduce the Courant factor to maintain stability. If one uses SD4B, the quality of the solution is already very good without the need to reduce the Courant factor, as in FV2L or SD4BL. It is therefore unclear which is the most efficient approach between doubling the resolution with FV2L or increasing the order with SD4B.

\begin{figure*}
    \centering
    \includegraphics[width=0.95\textwidth]{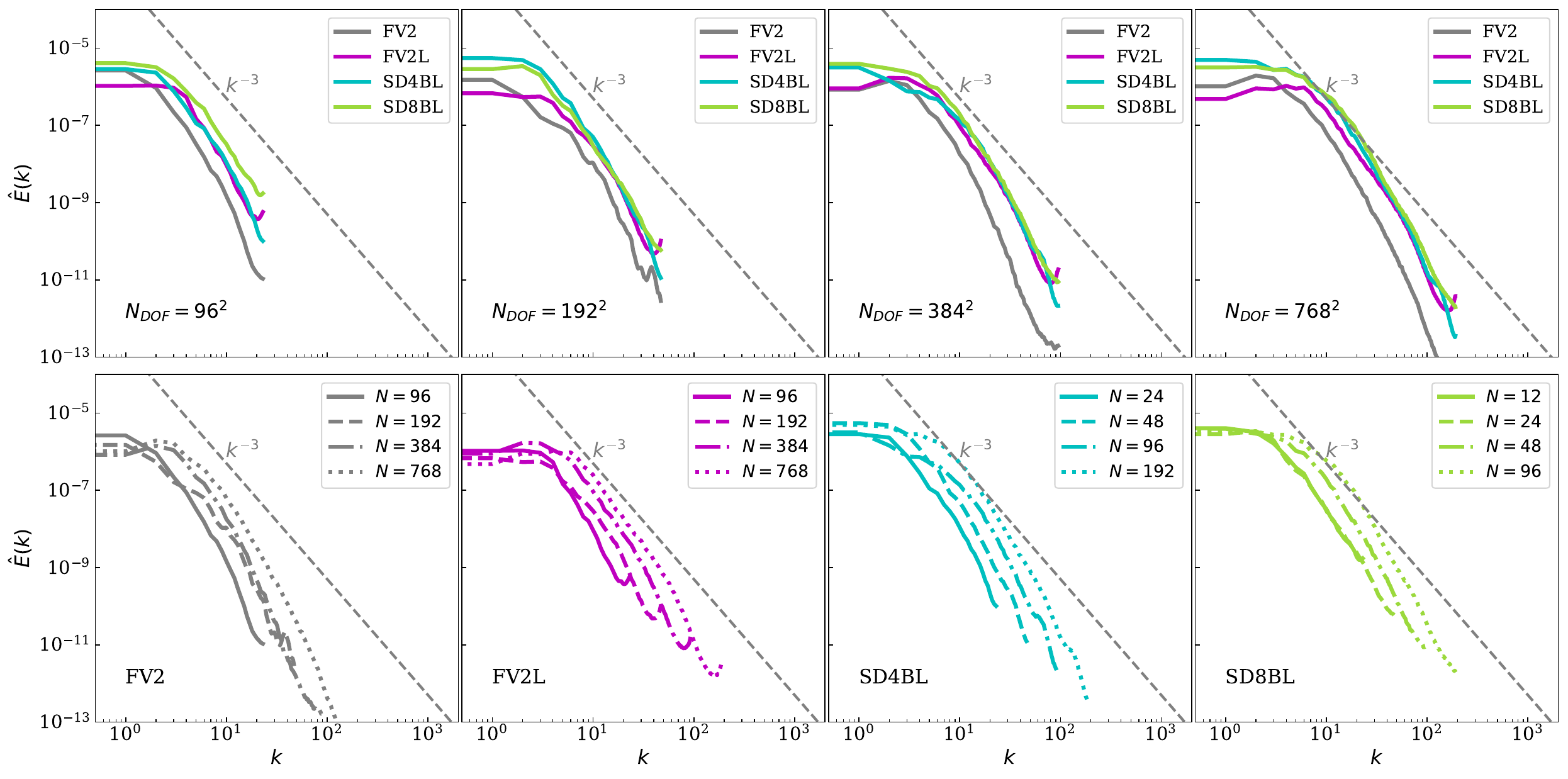}
    \caption{Kinetic energy power spectra $\hat{E}(k)$ at late time, averaged over 5 snapshots between time $t=1750$ and $t=2000$. The top row shows (from left to right) our different schemes using similar number of DOF. The bottom row on the other hand, shows the effect of increasing the number of DOF for a given numerical scheme. The dashed line shows the predicted scaling for 2D subsonic turbulence based on the conservation of enstrophy.}
    \label{fig:energy}
\end{figure*}

We show in \autoref{fig:energy} the kinetic energy power spectrum, in the convective zone, measured at late time, averaged over 5 snapshots between time $t=1750$ and $t=2000$. The figure follows the same conventions as Figure~\ref{fig:Ek_t}, expect that the x-axis is now the Fourier mode. For comparison, we show the theoretical scaling relation expected if one assumes perfect conservation of enstrophy \citep[see e.g.][]{Boffetta2010}. It is again striking to see how dissipative FV2 is compared to FV2L. Adding high-order polynomial reconstruction also helps preserving the power spectrum on large scales. On small scales, we see a steeper slope than the theoretical -3 scaling. When comparing to standard high-resolution spectral simulation of high-Reynolds number 2D turbulence \citep{Bracco2000}, we see the same steepening of the power spectrum at small scales. This is due to the formation of coherent structures at small scales such as the vortices seen in Figure~\ref{fig:bubbles_late}.  

In Figure~\ref{fig:energy}, the Nyqvist frequency is computed as $k_{\rm Nyq} = \pi/(p+1)N$, corresponding to the actual number of DOF for each method. This explains why all the power spectra corresponding to the same number of DOF in the top row of Figure~\ref{fig:energy} all reach the same maximum wavenumber. Although FV2 appears as the most dissipative scheme, FV2L competes quite well with the higher-order schemes SD4BL and SD8BL. Note the curious little upturn just before the Nyqvist frequency in case of FV2L. This might be an undesirable effect due to the low Mach number fix. SD4BL and SD8BL power spectra show a cleaner high-frequency tail, with SD8BL delivering the most power at the Nyqvist frequency.  Here again, we see that for this type of convective turbulence, using L-HLLC and using a high number of DOF seem to have a stronger effect than increasing the polynomial degree. The question of the actual cost of the different methods is therefore a central aspects of choosing one scheme over the others. 
From the present analysis, we can claim that our results with $p=3$ (SD4BL) and $p=7$ (SD8BL) are similar to those with FV2L at twice the resolution. 

It is important to estimate what fraction of the simulated volume is producing a truly high-order solution. For this, we monitor the fraction of troubled cells detected using PAD and NAD in our different simulations. The largest fraction {\it at late time} is around 25\% for our lowest resolution runs ($N_\text{DOF}=96$) and around 5\% for our highest resolution runs  ($N_\text{DOF}=768$). We always observed a burst of troubled cell detection at early time around 25\%. The late time fraction of troubled cells decreases systematically with resolution at fixed order. For example, for SD4B, we find the fractions to be 25\%, 15\%, 10\% and 5\% for $N=$24, 48, 96, 192. 

\section{Discussion}
\label{sec:discussion}

When evaluating the performance of our method for smooth solutions (Gresho vortex and hydrostatic equilibrium) we observe a clear and overwhelming advantage of using higher-order schemes in the low-Mach number regime. For more complex solutions involving discontinuities, the benefits of high-order become less obvious. We do observe more structure when going to high-order when performing the RTI test. This suggests that high-order methods provide higher effective Reynolds numbers than our reference second-order scheme.
Using a low-Mach number fix to the HLLC Riemann solver, we can reproduce the same lower numerical diffusivity increasing the spatial resolution of our second-order scheme.
This behavior is quite similar when studying the buoyantly rising bubble in an hydrostatic atmosphere. We see more small scale structures at high-order for a fixed number of DOF. 
It is difficult to evaluate what is the best between increasing the order or increasing the resolution at fixed order. The number of DOF used to model this type of problems seems to be the dominant factor.

Our results for the Gresho vortex are in agreement with those of \citet{nicoud2000conservative, desjardins2008high, klein2016high, Apsara, 2024arXiv240216706L}, finding that high-order methods are capable of evolving highly subsonic flows with almost negligible numerical viscosity. When using MUSCL-Hancock and HLLC we find significant degradation of the solution at lower Mach numbers, similar to \citet{2019ApJS..241...23D,DISPATCH}. We find that a Low-Mach number fix, like L-HLLC, is sufficient to enable Godunov-type second-order methods to capture these highly subsonic flows, recovering the same solution at all subsonic Mach numbers, as observed by \citet{2015A&A...576A..50M,2016arXiv161203910B,2019ApJ...875..128P}, though with more numerical viscosity when compared to high-order methods at the same $N_\text{DOF}$. We find that this difference is even more favorable for high-order methods when including a background velocity advecting the vortex. 

As observed in \citet{2021A&A...652A..53E,2022A&A...668A.143L}, we find that a well-balanced framework is necessary in order to properly capture the turbulent evolution of perturbations in the scenario of stellar convection. In \autoref{app:edelman} we include an additional test to validate our well-balanced implementation. 

Recent works have tested the L-HLLC Riemann solver when dealing with highly subsonic flows \citep{2019ApJ...875..128P,2022A&A...668A.143L,2024arXiv240216706L}, finding similar results than ours in terms of the improving the quality of the solution. Although the results with this fix are impressive at barely subsonic Mach numbers, it suffers from the same restriction on the CFL $\Delta t = \mathcal{O}(\mathcal{M}^2)$ as preconditioning methods \citep{birken2005stability,bruel2019low}. This restriction, when using explicit methods, becomes impossibly stringent as we reach the incompressible limit. It is important to stress that at high-order the low-Mach number fix is not strictly required, even with discontinuous solutions, as shown for the RTI and bubble tests. 

When analyzing the power spectrum of the kinetic energy, we do find a benefit of going to higher-order when compared to second-order. We observe more energy at small scales, a trend also observed when increasing the resolution. In this experiment, we do not observe a strong benefit of going from fourth- to eighth-order. The results for \nth{4} and \nth{8} order are similar to the ones obtained for \nth{2} order with $4\times$ more DOF, where the computational cost is similar for the \nth{4} order simulation and the \nth{2} order one with $4\times$ more DOF. Therefore, going to \nth{8} order in this case is counterproductive, because the simulation is more expensive than these other higher resolution simulations.

As mentioned before, the code used for this study is written in \texttt{python} using the \texttt{cupy} package. It should be considered as a proof of concept for our method, and its compatibility with GPU acceleration, before turning to a compiled language implementation. The algorithmic complexity of our method is proportional to $N_\text{DOF}=N_x(p+1)N_y(p+1)$, where the polynomial interpolation requires a stencil of $p+1$ points in each direction. Time integration adds a factor of $p+1$ for each time slice and $p+1$ Picard iterations. Therefore, the total complexity of the method is proportional to $N_xN_y(p+1)^5$ per time step. The Courant stability condition adds a factor of $N_x(p+1)$ to the overall cost to complete a simulation.
In the tests performed here, we have observed that before reaching the saturation of the GPU, the cost of doubling the resolution at fixed order of accuracy is of $2\times$ instead of $8\times$, the factor $2$ being solely due the increased number of time steps. As we increase the resolution more and more, we saturate the GPU and this factor goes back to the expected value of $8\times$. 

More interestingly, when we double the order of accuracy at fixed resolution, we should see a factor of $2^6\times$ increase in computational time. Before the GPU is fully saturated, we only see a factor of $2^2\times$ increase, due to the increase in the number of time steps and Picard iterations.  As we approach saturation of the GPU, this factor increase but only up to a factor of $2^5\times$. This suggests that in our experiments the GPU was able to absorb the factor of $2$ cost increase due to the polynomial interpolation. 
Note that it is not strictly necessary to use the same order of accuracy in time than the order of accuracy in space. We can use less time slices at the expense of reducing the Courant factor slightly to ensure stability. In doing so, the extra cost of moving from increasing the order (at the same $N_\text{DOF}$) can be reduced significantly, at the expense of losing accuracy in time.

\section{Conclusions}
\label{sec:conclusions}

In this work, we evaluate the performance of our implementation of the arbitrarily high-order Spectral Difference method, using the MUSCL-Hancock second-order Finite Volume (FV2) method as fallback scheme, when solving the compressible Euler equations for low Mach number flows. 

As shown in previous work, we observe that, for a classical second-order FV method (FV2), it is necessary to fix the Riemann solver (using for example L-HLLC) to properly reach the incompressible limit of the Euler equation at low Mach number. We believe that this fix is not necessary for our SD scheme, although it increases slightly the quality of the solution via our fallback scheme.

We observe that both methods need a well-balanced scheme in order to capture the physics of stellar convection at low Mach number but for different reasons.
For FV2, the well-balanced scheme is required to prevent truncation errors from dominating the numerical solution. For SD, the well-balanced framework is needed to improve the quality of the detection of troubled cells and therefore trigger the second-order fallback scheme appropriately. 

Our results highlight the amazing performance (exponential convergence) of high-order methods for smooth solutions, either for low Mach number flows or for small perturbations over hydrostatic equilibria. For simulations involving discontinuities, they do not show as clearly the added value of going to high-order. Our results suggest, nevertheless, that our high-order schemes outperform their low-order counterpart for the same, $2\times$ or even $4\times$ $N_\text{DOF}$. Another favorable aspect for high-order is that a low-Mach number fix is not strictly required, enabling explicit time-integration for this method. We observe in general that the cost-to-benefit of our best fourth-order scheme (SD4B) is better than all the other methods explored here.

\section*{Acknowledgements}
The simulations included in this work were executed on the Stellar cluster at Princeton University.

\section*{Data availability statement}
The data underlying this article will be shared on reasonable request to the corresponding author.


\bibliographystyle{mnras}
\bibliography{biblio} 



\appendix

\section{Performance}
\label{app:perf}
\begin{figure}
    \centering
    \includegraphics[width=.8\columnwidth]{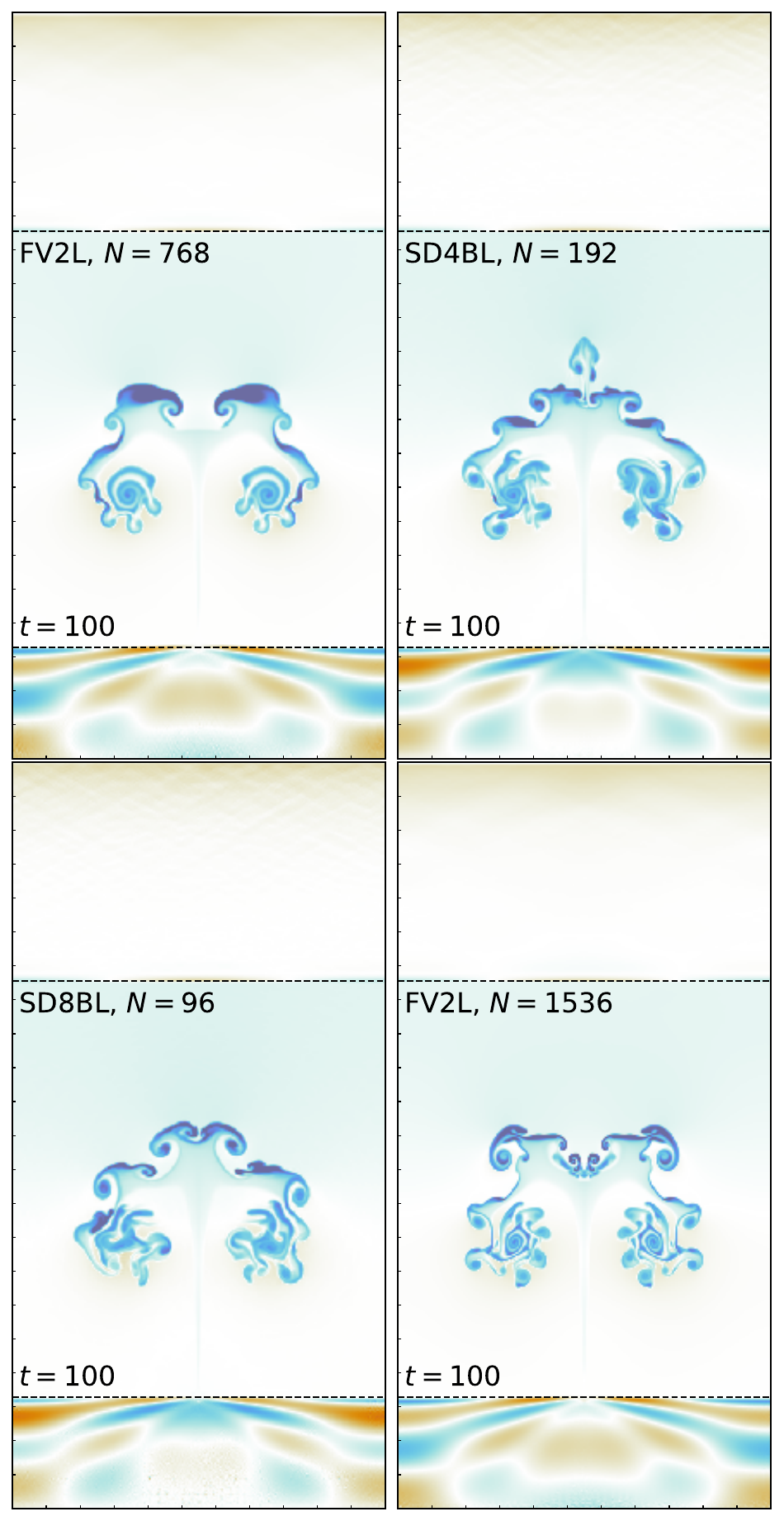}
    \caption{Temperature perturbation for the buoyantly rising bubble test at $t=100$. Results for FV2L, SD4BL and SD8BL with L-HLLC.}
    \label{fig:bubble-perf}
\end{figure}

Here we elaborate about the performance of the method. We omit presenting results for smooth solutions as the exponential convergence at high-order that vastly outperforms the low-order counterpart, is not representative of the behaviour observed with discontinuous solutions. In our experiments we find that the results at high-order are roughly equivalent to the ones at second-order for twice the resolution. In \autoref{fig:bubble-perf} we present snapshots for the bubble test, for the methods FV2L, SD4BL and SD8BL with $N_{DOF}=768$, and for FV2L with $N_{DOF}=1536$ (twice the resolution).

\begin{figure}
    \centering
    \includegraphics[width=\linewidth]{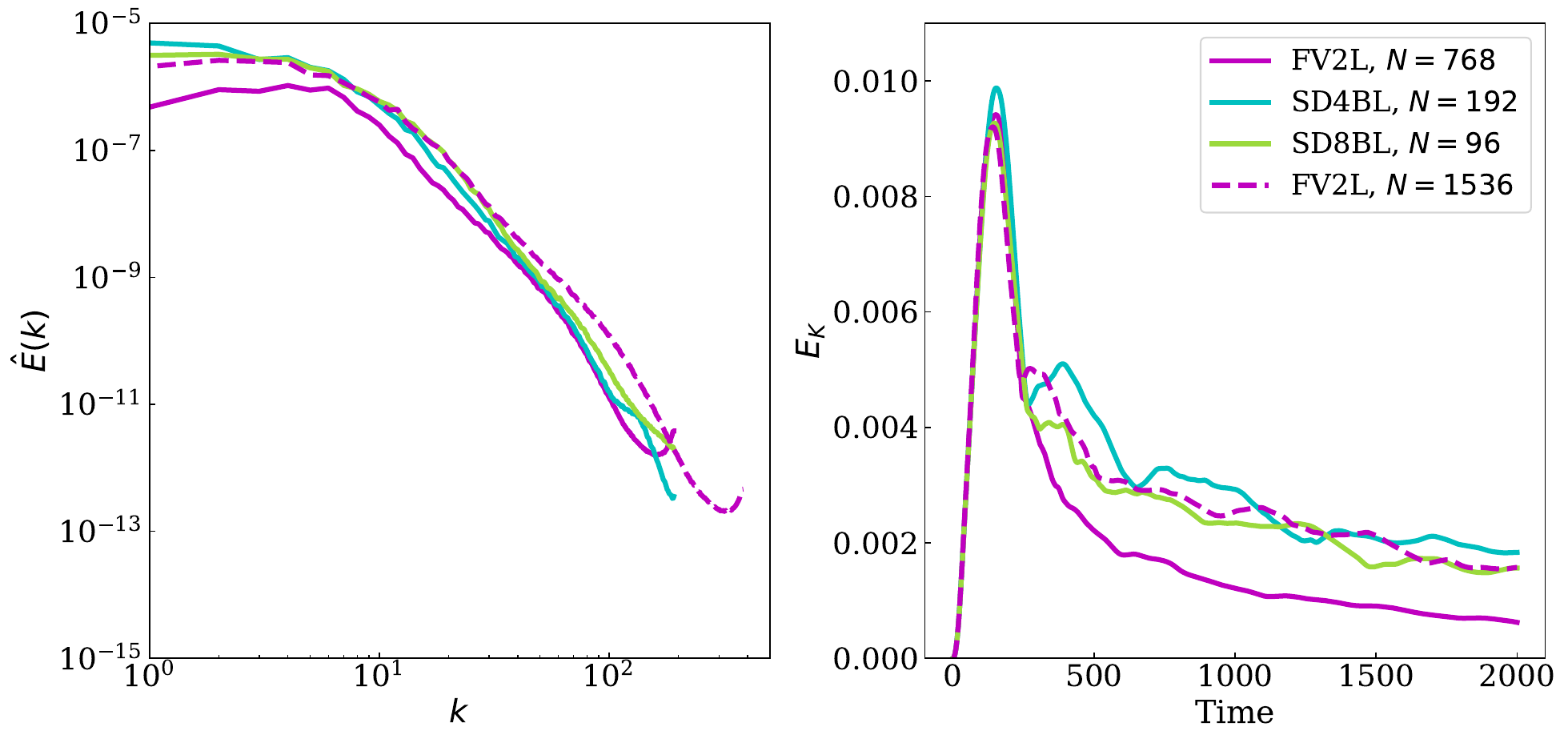}
    \caption{On the left the kinetic energy power spectra $\hat{E}(k)$ at late time, averaged over 5 snapshots between time $t=1750$ and $t=2000$. On the right the time evolution for the total kinetic energy of the buoyant bubble test for for FV2L, SD4BL and SD8BL with L-HLLC. }
    \label{fig:energy-perf}
\end{figure}

In \autoref{fig:energy-perf} we present the time evolution of the kinetic energy inside the convective region for the simulations shown in \autoref{fig:bubble-perf}. The energy conservation at low-order is similar to that obtained at high-order for half the resolution.

In \autoref{tab:cost} we present the computational cost for this set of simulations, the cost is normalized by the one obtained for FVL2 at $N_{DOF}=768$. This table shows that the cost for SD8BL is similar to that of FV2L at twice the resolution, whereas SD4BL takes half the cost. This leads us to the conclusion that, for discontinuous solutions, SD4BL emerges, between the methods studied here, as the optimal one in terms of cost to accuracy.
\begin{table}
    \centering
    \begin{tabular}{cccc}
        Computational cost  & FVL2 & SD4BL & SD8BL \\
        $N_{DOF}=768$   & 1x & 4.20x & 7.38x\\
        $N_{DOF}=1536$  & 7.75x & - & -\\
    \end{tabular}
    \caption{Comparison of the computational cost for the different methods used in this work. The cost is normalized to the time taken for FVL2 at $N_{DOF}=768$.}
    \label{tab:cost}
\end{table}

\section{Testing the stability of the low Mach number HLLC Riemann solver using a two-dimensional atmosphere}
\label{app:edelman}

Here we present an additional test to validate our well-balance scheme and to test the stability of the LHLLC Riemann solver. This test features a two-dimensional hydrostatic atmosphere and has been presented in \cite{2016A&A...587A..94K} and \cite{2021A&A...652A..53E}.
The hydrostatic solution is given by:
\begin{align*}
  \rho &= \rho_0\exp\left(-\frac{\rho_0}{P_0}\phi\right), \quad v_{x} = 0, \quad v_{y} = 0,\\
  P &= P_0\exp\left(-\frac{\rho_0}{P_0}\phi\right)
\end{align*}
where $\rho_0=1$, $P_0=1$ and $\phi(x) = gx$ is the gravitational potential corresponding with an acceleration of $g=1$ and $\gamma=5/3$. The computational domain is a box of dimensions $[0,1]\times[0,1]$ with $N_x = N_y = N$ elements in each direction. We use periodic boundary conditions in the $y$ direction. We impose the initial hydrostatic profile for ghost zones in the $x$ direction. 

\begin{figure}
\centering\includegraphics[width=\columnwidth]{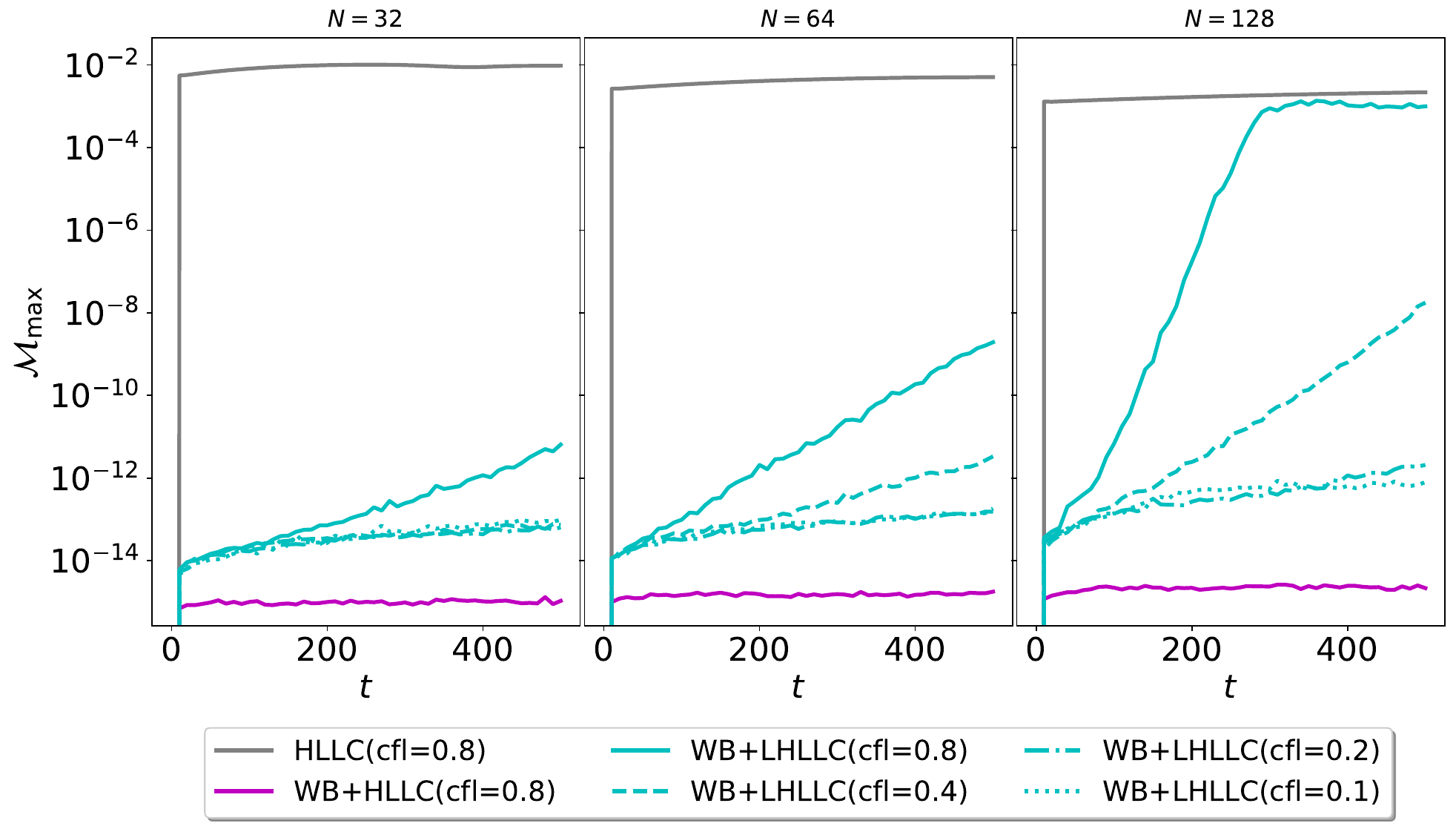}
    \caption{Two-dimensional isothermal hydrostatic atmosphere: time evolution of the maximum Mach number at different resolutions for the FV2 method with and without a well-balanced scheme, and with the HLLC and LHLLC Riemann solvers. For the LHLLC solver, different values for the CFL coefficient are used.}
    \label{fig:edelman}
\end{figure}

We used for this test the FV2 method with the standard HLLC Riemann solver. In \autoref{fig:edelman}, we present the time evolution for the maximum value of the Mach number on the grid. Similarly to the results shown in Figure~2 of \citet{2021A&A...652A..53E}, we observe that the FV2 scheme cannot preserve the hydrostatic profile without using a well-balanced scheme. We observe spurious fluctuations with ${\cal M}_{\rm max} \simeq 10^{-2}$, with the amplitude decreasing with resolution. We can see that our well-balanced scheme perfectly preserves the hydrostatic profile at all resolutions studied here. 

When we use the low-Mach number version of the Riemann solver (LHLLC), we observe the growth of an instability, also in striking agreement with the results shown in \citet{2021A&A...652A..53E}. We also observe that reducing the CFL coefficient helps reduce the growth rate of this instability (see the legends in Fig.~\ref{fig:edelman}). This experiment helps us adopting the proper Courant condition for the simulations performed in this paper. 

We can interpret this effect using a simplified modified equation analysis using the advection equation. We have
\begin{equation}
\frac{\partial u}{\partial t} + a \frac{\partial u}{\partial x}=0
\end{equation}
Here, $a$ is the advection velocity but it can be interpreted as $\pm c_s$ (plus or minus the sound speed) and $u$ as the corresponding eigenvector amplitude for the linearized Euler equations.
This equation is discretized in one dimension using the standard Godunov approach:
\begin{equation}
    \frac{u^{n+1}_i - u^n_i}{\Delta t} + \frac{f^{n+1/2}_{i+1/2}-f^{n+1/2}_{i-1/2}}{h} = 0
\end{equation}
The Godunov flux is obtained using the upwind solution as
\begin{equation}
    f^{n+1/2}_{i+1/2} = a \frac{u^n_{i} + u^n_{i+1}}{2} - \left|a\right| \frac{u^n_{i+1} - u^n_{i}}{2}
\end{equation}
In deriving the low-Mach number version of HLLC, we multiply the diffusive term by the local Mach number ${\cal M} = \left|v\right| / a$ so that the Godunov flux becomes
\begin{equation}
    f^{n+1/2}_{i+1/2} = a \frac{u^n_{i} + u^n_{i+1}}{2} - \left|a\right| {\cal M} \frac{u^n_{i+1} - u^n_{i}}{2}
\end{equation}
We can now Taylor-expand the resulting numerical scheme both in time and space. We get to leading order in $h^2$:
\begin{equation}
    \frac{\partial u}{\partial t} + a \frac{\partial u}{\partial x} = - \frac{\Delta t}{2} \frac{\partial^2 u}{\partial t^2} + \frac{\left|a\right|h}{2}{\cal M} \frac{\partial^2 u}{\partial t^2}
\end{equation}
Exploiting the original equation with $\frac{\partial^2 u}{\partial t^2}=a^2 \frac{\partial^2 u}{\partial x^2}$, we finally get the classical result:
\begin{equation}
    \frac{\partial u}{\partial t} + a \frac{\partial u}{\partial x} = \nu \frac{\partial^2 u}{\partial x^2} 
\end{equation}
where $\nu \simeq \left|a\right|h \left( {\cal M} - C\right)$ is the numerical diffusion coefficient and $C=\left|a\right| \Delta t/h$ is the Courant factor. If ${\cal M}=1$, we recover the classical Courant stability condition that states that $\nu>0$ only if $C<1$. However, for LHLLC we have a much more restrictive condition, namely $C<{\cal M}$. Note that this low-Mach number stability condition for LHLLC is similar to the stability condition for time-explicit preconditioning techniques for low Mach number flow \citep{Birken2005}.

If one does not satisfy the Courant stability condition, the numerical solution will develop instabilities due to a negative diffusion coefficient with growth rate
\begin{equation}
    \Gamma = -\nu k^2 \simeq C \left| a \right| h k^2 \propto \frac{C\left| a \right| }{h} 
\end{equation}
where $k$ is the wavenumber of the largest Fourier mode on the grid and we have neglected in the previous equation the very small values of ${\cal M}\leq 10^{-2}$ compared to $C \geq 0.1$ in the numerical experiment we discuss here.  The analytical form we found for the instability growth rate agrees qualitatively with our numerical results, with the correct dependence with respect to the Courant factor and the grid resolution. Note that this agreement can be explained by the fact that these small-amplitude sound waves are almost perfectly linear, which aligns nicely with our simplified analysis. We conclude that our implementation of LHLLC is unstable unless one uses a very restrictive Courant condition $C < {\cal M}$, and that we can delay the growth of the instability by reducing the Courant factor, especially at high resolution.


\bsp	
\label{lastpage}
\end{document}